\newif\ifarxiv
\arxivtrue 

 \documentclass[
 reprint,
 amsmath,amssymb,
 aps,
]{revtex4-1}
\newcommand{\smsec}[1]{Supplemental Material \cite[Sec.~{#1}]{supp}}

\ifarxiv

\newcommand{\revisedversion}[1]{{\color{black} #1}}
\else
\newcommand{\revisedversion}[1]{{\color{blue} #1}}
\fi

\usepackage{hyperref}
\usepackage{graphicx} 
\usepackage{dcolumn} 
\usepackage{bm} 
\usepackage{subcaption} 
\usepackage{soul}
\usepackage{xcolor} 
\usepackage{multirow}
\usepackage[version=3]{mhchem}
\usepackage{url}
\usepackage{amssymb}
\usepackage{comment} 

\ifarxiv 
	\graphicspath{{./}}
\else 
	\graphicspath{{z_gene/}}
\fi

\begin{document}

\title{Using transcription-based detectors to emulate the behaviour of sequential probability ratio-based  concentration detectors}

\author{Chun Tung Chou}
\email{c.t.chou@unsw.edu.au}
\affiliation{%
School of Computer Science and Engineering, University of New South Wales, Sydney, NSW 2052, Australia
}


\begin{abstract}
The sequential probability ratio test (SPRT) from statistics is known to have the least mean decision time compared to other sequential or fixed-time tests for given error rates. In some circumstances, cells need to make decisions accurately and quickly, therefore it has been suggested the SPRT may be used to understand the speed-accuracy tradeoff in cellular decision making. It is generally thought that in order for cells to make use of the SPRT, it is necessary to find biochemical circuits that can compute the log-likelihood ratio needed for the SPRT. However, this paper takes a different approach. We recognise that the high-level behaviour of the SPRT is defined by its positive detection or hit rate, and the computation of the log-likelihood ratio is just one way to realise this behaviour. In this paper, we will present a method which uses a transcription-based detector to emulate the hit rate of the SPRT without computing the exact log-likelihood ratio. We consider the problem of using a promoter with multiple binding sites to accurately and quickly detect whether the concentration of a transcription factor is above a target level. We show that it is possible to find binding and unbinding rates of the transcription factor to the promoter's binding sites so that the probability that the amount of mRNA produced will be higher than a threshold is approximately equal to the hit rate of the SPRT detector. Moreover, we show that the average time that this transcription-based detector needs to make a positive detection is less than or equal to that of the SPRT for a wide range of concentrations. We remark that the last statement does not contradict Wald's optimality result because our transcription-based detector uses an open-ended test. 
\end{abstract}


\maketitle

\section{Introduction}
\label{sec:intro}

Cells often need to detect whether the concentration of a particular chemical species is above or below a target level. In some circumstances, e.g. in embryo development, it is important that cells can do this detection task both accurately and quickly. From a mathematics point of view, fast and accurate detection is the key goal of Wald's Sequential Probability Ratio Test (SPRT) \cite{Wald}. In the SPRT, the log-likelihood ratio of a sequence of observations is used to decide between two hypotheses. Wald and Wolfowitz showed that, for given false positive and false negative error rates, the SPRT requires the least mean number of observations compared to other sequential or fixed-time tests \cite{Wald.1948}. The paper \cite{Siggia:2013dd} is the first to suggest to use the SPRT to understand the speed-accuracy tradeoff in cellular decision making. Recently, \cite{Desponds:2020bla} showed that the SPRT is a possible strategy that can enable the nuclei in {\sl Drosophila melanogaster} (fruit fly) embryos to accurately and quickly detect the level of the morphogen Bicoid. Given that fast and accurate concentration detection is also a requirement in many areas of cellular operations \cite{Siggia:2013dd}\cite{Aquino:2015em}\cite{Granados:2017ig}, it is therefore important to understand how biochemical circuits can be used to enable fast and accurate concentration detection. \\

There is few work \cite{Siggia:2013dd,Desponds:2020bla} on studying how biochemical circuits can be used to realise the SPRT. Both \cite{Siggia:2013dd,Desponds:2020bla} took a direct approach where their focus is on finding biochemical circuits that can approximately compute the log-likelihood ratio, e.g.~\cite{Siggia:2013dd} used a protein circuit while \cite{Desponds:2020bla} used gene transcription coupled with mRNA degradation. We take a different approach in this paper. We see the goal of the SPRT is to achieve a set of high level behaviours, e.g. hit rate, mean decision time etc. Our view is that the computation of the log-likelihood ratio is only a means to an end (= realising the high level behaviour), not an end to itself. In this paper, we present a transcription-based detector whose high level behaviour is similar to
that of the SPRT detector, and we achieve that without computing the exact log-likelihood ratio. \\

This paper considers a concentration detection problem whose goal is to detect whether the concentration of a specific chemical species is above a target level in an accurate and fast manner. We do this by embedding this concentration detection problem within the SPRT framework where the null (resp. alternative) hypothesis corresponds to a low (high) concentration. The standard SPRT uses two boundaries to decide between the hypotheses. If the log-likelihood ratio of the SPRT hits the upper boundary first, then the SPRT will decide that the concentration is above the target level. The high level behaviour of this SPRT detector can be characterised by the probability that the log-likelihood ratio will hit the upper boundary and we will refer to this as the hit rate of the SPRT detector. In this paper, we consider the problem of using a promoter with multiple binding sites to accurately and quickly detect whether the concentration of a transcription factor is above a target level. We will show that it is possible to find binding and unbinding rates of the transcription factor to the binding sites so that the probability that the amount of mRNA transcribed will hit a boundary level is approximately equal to the hit rate of the SPRT detector. Moreover, we show that the average time that this transcription-based detector needs to make a positive decision is less than or equal to that of the SPRT for a wide range of transcription factor concentrations. We remark that the last statement does not contradict SPRT's optimality result because our transcription-based detector uses an open-ended test \cite{BassevilleN}. The main result is that we are able to use the transcription-based detector to emulate the hit rate of the SPRT.   \\

The rest of the paper is organised as follows. Section \ref{sec:sprt} defines the concentration detection problem in the context of this paper and recalls some key result of the SPRT detector. Section \ref{sec:gene} describes our proposed transcription-based detector. 

\section{Concentration detection using the SPRT}
\label{sec:sprt}
This section describes how the SPRT can be used for concentration detection with minimal mean decision time. This type of SPRT problems is first studied in \cite{Siggia:2013dd}. We have adapted the problem description to fit the context of this paper. \\

The detection problem is defined using a reaction pathway which consists of a gene promoter with multiple binding sites and a transcription factor \ce{L}. We assume that the promoter has $m$ (where $m$ = 1, 2, ... ) copies of the same type of binding site \ce{X}. Each site \ce{X} can exist in two states: unbound \ce{X} and bound \ce{X_*}. We assume that the binding and unbinding reactions are modelled by:
\begin{subequations}
\label{cr:g:u0:all} 
\begin{align}
\cee{
X + L &  ->[g_+] X_* + L \label{cr:g:u0:1}  \\
X_* &  ->[g_-] X \label{cr:g:u02} 
} 
\end{align}
\end{subequations}
where $g_+$ and $g_-$ are reaction rate constants. We assume that the binding of \ce{L} to each of the $m$ binding sites of type \ce{X} is probabilistically independent of each other.
For the SPRT, we assume that the concentration of \ce{L} does not change over time and use $L$ to denote this concentration. In addition, we model the reactions \eqref{cr:g:u0:all} using the Chemical Master Equation (CME) \cite{Gardiner}. This means that each site \ce{X} switches between the unbound and bound states in a stochastic manner. The aim of the detection problem is to use the binding history of the site \ce{X} to infer information on the concentration $L$. \\


We have now defined the reaction pathway and its model. In order to define the detection problem, we will need to specify the measured data and the hypotheses. The measured datum $X_*(t)$ at time $t$ is the number of type X sites that are bound, hence $X_*(t)$ is an integer in the interval $[0,m]$; e.g., $X_*(t) = 0$ means that none of the \ce{X} sites is bound at time $t$. 
Since the SPRT is based on a sequence of measurements, we assume that at time $t$, the data available to the detection problem are the continuous-time history of $X_*(\tau)$ for all $\tau \in [0,t]$; we will use ${\cal X}_*(t)$ to denote this history. The aim of the detection problem is to use the measured data ${\cal X}_*(t)$ to decide whether the concentration $L$ (which produces the data ${\cal X}_*(t)$) is at $L_0$ or $L_1$ where $L_0$ and $ L_1$ are two given reference concentration levels with $L_0 < L_1$. In other words, the hypotheses are $L$ is $L_0$ or $L$ is $L_1$. The decision of the SPRT is based on computing the log-likelihood ratio $R(t)$:
\begin{eqnarray}
R(t) = \log \left( \frac{ {\rm Pr}[ {\cal X}_*(t) | L_1  ] }{  {\rm Pr}[ {\cal X}_*(t) | L_0] }   \right) \label{eq:llr} 
\end{eqnarray}
where ${\rm Pr}[ {\cal X}_*(t) | L_i]$ is the conditional probability of observing the history ${\cal X}_*(t)$ assuming that the concentration of the transcription factor is $L_i$ ($i = 0, 1$). Note that the concentration $L$ corresponds to the actual concentration that generates the data ${\cal X}_*(t)$ and is not limited to $L_0$ or $L_1$. Later on, we will consider the property of the detector over a range of $L$. \\

The stochastic properties of the log-likelihood ratio $R(t)$ 
can be well modelled by a Wiener process with two parameters: drift and diffusivity. The drift of $R(t)$ is related to the rate of change of the mean of log-likelihood ratio:
\begin{eqnarray}
\frac{d{\rm E}[R(t)] }{dt} & = & m
\underbrace{
\frac{g_+ g_- }{g_- + g_+ L} \left(  L \; \log \left(\frac{L_1}{L_0} \right) - (L_1 - L_0)  \right)}_{V} 
\nonumber \\  \label{eq:llr:v} 
\end{eqnarray}
where ${\rm E[ \; ]}$ denotes expectation. Eq.~\ref{eq:llr:v} says that the drift of $R(t)$ is $m \cdot V$ where $V$ is the drift of $R(t)$ when there is only one type X binding site (i.e., $m = 1$). (The proof of \eqref{eq:llr:v} is given in \smsec{\ref{app:proof:llr:v}}.)
The diffusivity of $R(t)$ is \revisedversion{proportional to}  the rate of change of the variance of $R(t)$, and is of the form $m \cdot D$ where $D$ is the diffusivity of $R(t)$ when $m = 1$. 
 An expression of $D$ in terms of the parameters of the detection problem can be found in \cite{Siggia:2013dd}. Since the expression of $D$ is fairly long, we have included it as \eqref{app:eq:llr:D} in \smsec{\ref{app:llr:diff}}. Furthermore, \smsec{\ref{app:llr:diff}} explains how the expression of the $D$ can be derived. We remark that both the drift and diffusivity of $R(t)$ are proportional to $m$ because the log-likelihood ratio $R(t)$ can be written as a sum of $m$ independent and identically distributed random variables since the binding of \ce{L} to the $m$ sites is assumed to be independent. 

Note that both $V$ and $D$ are functions of the concentration $L$ but we will only indicate this dependence when it is necessary.  \\


The decision in the SPRT is made by using the log-likelihood ratio $R(t)$ in conjunction with two boundaries $K_-$ and $K_+$ where $K_- < 0 < K_+$. We assume that the initial log-likelihood ratio $R(0) = 0$. If for all $\tau \leq t$, we have $K_- < R(\tau) < K_+$, then the SPRT detector is in an undecided state at time $t$. If, on the other hand, the log-likelihood ratio $R(t)$ hits the boundary $K_+$ (resp. $K_-$) first, then SPRT decides that the hypothesis $L_1$ ($L_0$) holds. These boundaries determine the error rates for the detection problem \cite{Wald}. In this paper, we restrict ourselves to the equal error case where both $K_+$ and $-K_-$ are equal to a constant $K (> 0)$. The performance of the SPRT concentration detector can be determined by studying the first passage time that the aforementioned Wiener process will hit a boundary.  In this paper, our concern is to detect whether the concentration $L$ is above a target level, so we will only consider whether $R(t)$ hits the upper boundary $K$. The probability $H$ (or hit rate) that the log-likelihood ratio $R(t)$ will hit the upper boundary is:
\begin{eqnarray}
H & = & \frac{1}{2}  + \frac{1}{2}  \tanh \left( \frac{V K}{2 D} \right) \label{eq:llr:h} 
\end{eqnarray}
Note that the expression in \eqref{eq:llr:h} is equivalent to that in \cite{Siggia:2013dd}. We have written it in this particular form so that we can later on relate it to our proposed transcription-based detector. We remark that the hit rate $H$ is independent of the value of $m$.
Let $L_{0.5}$ be the concentration $L$ such that $V(L_{0.5}) = 0$, and consequently $H(L_{0.5}) = 0.5$. From \eqref{eq:llr:v}, we have $L_{0.5} = (L_1 - L_0) / \log \left( \frac{L_1}{L_0}  \right)$. We can see from \eqref{eq:llr:v} and \eqref{eq:llr:h} that the probability of deciding for hypothesis $L_1$ is greater than half if $L > L_{0.5}$.  We can therefore use the SPRT detector to decide whether the concentration $L$ is above the target level $L_{0.5}$. Note that if the reference concentrations $L_0$ and $L_1$ are close to each other, then $L_{0.5} \approx \frac{L_0+L_1}{2}$. \\

For all the trajectories of the log-likelihood ratio $R(t)$ that hit the upper boundary, the mean time $F$ to decide for hypothesis $L_1$, is given by: 
\begin{eqnarray}
F & = & \frac{K}{mV} \tanh \left( \frac{V K}{2 D} \right) \label{eq:llr_fpt}.
\end{eqnarray}
This result does not appear to be well known. This is because most analyses on the SPRT focused on computing the weighted mean time to reach the two boundaries rather than the time to reach a specific boundary. We have included a derivation in \smsec{\ref{app:llr_fpt_upper}}. We remark that \eqref{eq:llr_fpt} says that we can decrease the mean decision time $F$ by using a higher value of $m$.
Note that both $H$ and $F$ are dependent on the concentration $L$ and we will indicate that when necessary. \\


The advantage of using SPRT is that for given error rates, the SPRT has the least mean decision time compared to other sequential or fixed-time tests provided that the mean decision time is finite \cite{Wald.1948}. 
This is the reason why we have chosen the emulate the SPRT in order to achieve fast concentration detection. 
Note that the expression of the least mean decision time is given in \eqref{eq:llr_fpt}.
\\

Although we have used both hit rate $H$ and mean first passage time $F$ to characterise the behaviour of the SPRT, we need to point out $H$ are $F$ are not independent of each other. For example, we can determine $F$ from $H$, see \cite{Cox,BassevilleN}. In this paper, we will focus on imitating the hit rate $H$ of the SPRT. 


\section{Gene promoter for concentration detection}
\label{sec:gene}
\begin{figure}[t]
        \centering
        \includegraphics[page=1,scale = 0.6,trim={20 400 450 100},clip]{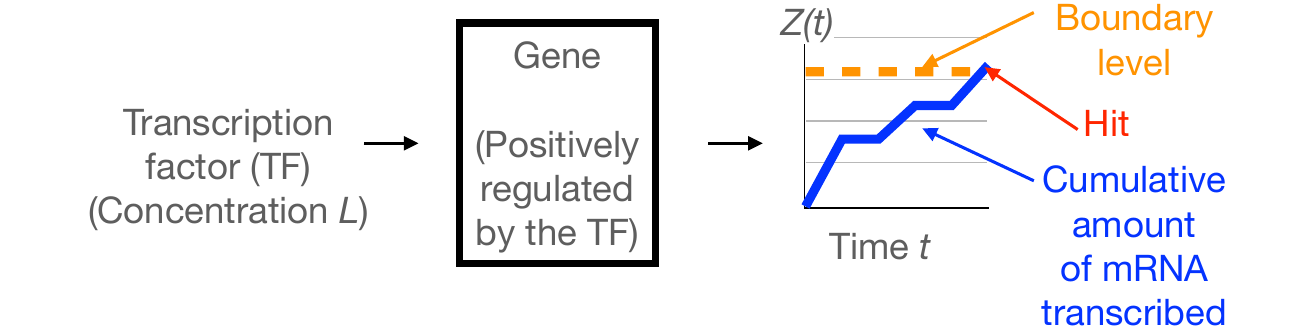} 
\caption{Transcription-based detector. 
}      
\label{fig:trans_detector}
\end{figure}

This section will present a transcription-based detector whose aim is to detect whether the concentration $L$ of the transcription factor is above a target level. 
This detector, which is depicted Fig.~\ref{fig:trans_detector}, uses a gene which is positively regulated by the transcription factor \ce{L}. The detector uses the cumulative amount of mRNA transcribed $Z(t)$ and a positive boundary level to make a decision. We assume $Z(0) = 0$. If $Z(\tau)$ is less than the boundary level for all $\tau \leq t$, then the detector is at an undecided state at time $t$. Otherwise, if $Z(t)$ hits the boundary level for some time $t$, then the detector decides that a hit has occurred. Our aim is to derive a transcription-based detector whose hit rate is approximately equal to the hit rate $H$ of the SPRT detector in \eqref{eq:llr:h}. We can classify our detector as an open-ended sequential detector \cite{BassevilleN} which uses only one boundary rather than two boundaries as in SPRT. Note that our detector has only two possible states: decided that $L$ is above the target level or undecided; in other words, the detector never decides that the $L$ is below the target level. Biologically, this means the cell reacts when the concentration $L$ is above the target level and does nothing otherwise. \\

We will divide the derivation into two parts. In Sec.~\ref{sec:gene:mean}, we derive a class of gene promoters whose mean transcription rate is approximately equal to the mean log-likelihood ratio ${\rm E}[ R(t) ]$ in \eqref{eq:llr:v} for sufficiently large $L$. After that, in Sec.~\ref{sec:gene:hit} we show how to use the amount of mRNA transcribed for concentration detection. 

\subsection{A gene promoter that can approximately compute the mean log-likelihood ratio}
\label{sec:gene:mean}
Since we are only interested in deciding whether a concentration is above a target level, we will focus on the mean log-likelihood in \eqref{eq:llr:v} when it is positive. Consider the following differential equation: 
\begin{eqnarray}
\frac{ dM(t)}{dt}  & = &  
\underbrace{ g_- \;  \log \left(  \frac{L_1}{L_0} \right) }_r 
\; m \bar{X}_*   \; 
\underbrace{\left[  1 -  \frac{L_{0.5}}{ L}  \right]_+ }_{Q} 
\label{eq:gene:v} 
\end{eqnarray}
where $[w]_+ = \max(w,0)$ and $\bar{X}_* = \frac{g_+ L}{g_- + g_+ L}$ is the mean fraction of time that a site \ce{X} is bound. It can be shown that if $L > L_{0.5}$, then the right-hand sides of \eqref{eq:llr:v} and \eqref{eq:gene:v} are equal. Furthermore, if ${\rm E}[R(0)] = M(0) = 0$, then $M(t) = E[R(t)]$ for $L > L_{0.5}$. 
The derivation of \eqref{eq:gene:v} can be found in \smsec{\ref{app:proof:gene:v}}.  \\
 
The importance of \eqref{eq:gene:v} is that it can be approximately realised by gene transcription. We will first consider the case when $m = 1$ and our goal is to find a promoter whose mean transcription rate is approximately equal to $r \bar{X}_* Q$, which is the right-hand side of \eqref{eq:gene:v} for $m = 1$. Note that the threshold-hyperbolic function $Q$ in \eqref{eq:gene:v} has the property that $0 \leq Q \leq 1$. We will interpret $\bar{X}_* Q$ as the probability that a promoter is active and $r$ as the mean transcription rate when the promoter is active. According to the thermodynamic model of gene transcription \cite{Ackers:82,Phillips:2019hj}, we can interpret $M(t)$ as the mean amount of mRNA transcribed by time $t$ assuming that $M(0) = 0$. \\
 
The question now is how we can find a promoter so that the probability that it is active is given by $\bar{X}_* Q$. We consider a promoter that has multiple binding sites that can be bound by the transcription factor \ce{L}. We divide these binding sites into two groups. The first group consists of only one binding site and this site behaves in the same way as \ce{X} in Sec.~\ref{sec:sprt}. We will use \ce{X} to refer to this binding site and the probability that this site is bound is $\bar{X}_*$. We will use \ce{Y} to denote the second group of binding sites. We will show shortly that there are binding and unbinding rates so that the probability that all the sites in \ce{Y} are bound is approximately equal to $Q$. We assume that the promoter (denoted by \ce{X-Y}) is active when all its binding sites are bound. If the binding of \ce{L} to \ce{X} is independent of the binding of \ce{L} to the sites in \ce{Y}, then the probability that the promoter is active is then $\bar{X}_* Q$. Note that the last statement can be formally justified using the Product Theorem in \cite{Ahsendorf:2014fc}. Next we will explain how we can approximately realise the threshold-hyperbolic function $Q$. \\

It was proved in \cite{Gunawardena:2005jm} that, if $|\lambda u| > 1$, then 
\begin{eqnarray}
\lim_{n \rightarrow \infty} 
\underbrace{
\frac{ (\lambda u)^n }{1 + \lambda u + ... +(\lambda u)^n }
}_{f_n(u)} & = & 1 - \frac{1}{\lambda u}
\label{eq:thf:approx}
\end{eqnarray}
The proof is based on convergence of geometric series. This means that we can approximately realise $Q$ in \eqref{eq:gene:v} as a function of $L$ by using using the rational function $f_n(L)$ for some sufficiently large $n$ with $\lambda =  \frac{1}{L_{0.5}}$. We next explain how $f_n(L)$ can be realised by using $n$ binding sites in \ce{Y}; our method uses the linear framework for gene transcription in \cite{Ahsendorf:2014fc,Estrada:2016ct}.  \\ 

Since \ce{Y} has $n$ binding sites, there are $2^n$ possible microstates for \ce{Y} where each microstate is defined by whether the sites are bound by \ce{L} or not. We assume that the binding rate of \ce{L} to a microstate and the unbinding rate from a microstate depend only on the number of sites in the microstate that has been bound. Let \ce{Y_j} (where $j = 0, 1, \ldots, n$) denote the macrostate of \ce{Y} that has $j$ sites bound. We can interpret \ce{Y_j}'s as the states of a continuous-time Markov chain (CTMC) and write its state-transition diagram as:
\begin{align}
\cee{
Y_0 &  <=>[a_1 L][b_1]  Y_1 <=>[a_2 L][b_2] Y_2 \; ... \; Y_{n-1}  <=>[a_n L][b_n] Y_n  
} 
\label{eq:g:y}
\end{align}
where $a_1 L$, $b_1$ etc. are the state transition rates. If we choose $a_1,b_1 ...$ such that $\frac{a_i}{b_i} = \lambda$ for $=1,...,n$, then it can be verified that the probability that \ce{Y} is in state \ce{Y_n} is equal to $f_n(L)$, see \smsec{\ref{app:Yn}}. In other words, $Q$ in \eqref{eq:gene:v} is approximately equal to the probability that all sites in \ce{Y} are bound. This explains how $Q$ in \eqref{eq:gene:v} can be realised. Note that we have only used half of the $2n$ degrees-of-freedom (i.e., half of the parameters among $a_i$, $b_i$) and we will use the rest to control the fit to the hit rate later on. We remark that the above derivation corresponds to the case where the promoter \ce{X-Y} is in thermal equilibrium with the transcription factor \ce{L} because detailed balance holds, see \cite{Estrada:2016ct}. This completes the description for the case where $m = 1$.
\\

In general, when there are $m$ independent and identical binding sites of type X, the corresponding transcription-based detector should have a mean transcription rate of $r m \bar{X}_*  Q$ according to \eqref{eq:gene:v}. In this case, the promoter has the form X$_{\rm m}$--Y with a total of $m+n$ binding sites where X$_{\rm m}$ denotes the $m$ binding sites of type X and the $n$ binding sites in Y are used to implement the threshold-hyperbolic function $Q$ as before. Furthermore, the promoter should behave as follows:
\begin{enumerate}
\item In the microstate where exactly $k$ of type X sites are bound and all sites of Y are bound, the mean transcription rate is $kr$ where $k = 1, \ldots, m$. 
\item No transcriptions occur in all other microstates. 
\end{enumerate}
We show in \smsec{\ref{app:microstates}} that these transcription rules will give the mean transcription rate $r m \bar{X}_*  Q$.  \\

We remark that in our previous work in \cite{Chou:2018jh,Chou:2019gf,Chou.2021}, we realised the threshold-hyperbolic function by fitting it to a Hill function. Many papers, e.g. \cite{Phillips:2019hj,Estrada:2016ct}, have pointed out that Hill functions are phenomenological model and do not reflect thermodynamic reality. The model \eqref{eq:g:y} is based thermodynamic microstate \cite{Estrada:2016ct} and does not suffer from this problem.  

\subsection{Imitating the hit rate}
\label{sec:gene:hit}
In this section, we will use the concentration of the mRNA molecules transcribed by the gene \ce{X_{\rm m}-Y} together with the boundary $K$ to realise a detector whose hit rate is approximately equal to the hit rate $H$ of the SPRT in \eqref{eq:llr:h}. Let $Z(t)$ denote the amount of mRNA transcribed by \ce{X_{\rm m}-Y} up to time $t$. We will first explain how $Z(t)$ can be modelled. The first method that we will use to model $Z(t)$ is via a CME and this will be used in simulation. The promoter \ce{X_{\rm m}-Y} has $2^{m+n}$ microstates and let $P_j(t)$ be the probability that the promoter is in the microstate indexed by $j$ \revisedversion{at time $t$.} Let also $r_j$ be the mean transcription rate when the promoter is microstate $j$. It follows that $P_j(t)$ can be obtained as a realisation of the CTMC that describes the evolution of the microstates of the promoter \ce{X_{\rm m}-Y}. We will obtain $Z(t)$ by $ \int_{0}^t \sum_{j} r_j P_j(\tau) d\tau$.
However, it is difficult to analyse this model of $Z(t)$. So, in the second method, we model $Z(t)$ approximately by a Wiener process with drift $\widetilde{V}(L)$ and diffusivity $\widetilde{D}(L)$. We explain how $\widetilde{V}$ and $\widetilde{D}$ can be computed using the binding and unbinding parameters in \smsec{\ref{app:promoter_VD}}. \\ 

Our transcription-based detector decides that the concentration $L$ is above the target level if $Z(t)$ hits the boundary $K$ for some time $t$. (Note that the transcription-based detector and the SPRT use the same boundary $K$.) If the drift $\widetilde{V} > 0$, we know from \cite{Cox} that the probability that $Z(t)$ hitting the boundary $K$ is 1. This certainly does not allow us to emulate the behaviour of SPRT. We resolve this by assuming that the transcription factor \ce{L} is present only for a finite duration $T$. Note that for SPRT, we still assume that the concentration $L$ lasts for an infinite duration.  We remark that these are not incompatible assumptions since SPRT represents the ideal behaviour that we want to imitate while mRNA production represents a practical realisation, so it is legitimate to have infinite signal duration for one case and finite signal duration for the other. We further assume that that the transcription of \ce{X_{\rm m}-Y} ceases at time $T$ when the transcription factor signal ends; alternatively, the cessation of transcription can also be caused by the onset of mitosis \cite{Lucas.2018}. For this set up, if the trajectory $Z(t)$ hits the boundary $K$ for some time $t \leq T$, then the transcription-based detector decides that the concentration $L$ is above the target level and this is counted as a hit; otherwise the detector remains in an undecided state. We assume that the duration $T$ is shorter than the mRNA degradation time-scale so that we can neglect mRNA degradation for $t \leq T$. This means $Z(t)$ is non-decreasing for $t \leq T$. \\

We show in \smsec{\ref{app:promoter_fpt_upper}} that the hit rate $\widetilde{H}$ of the transcription-based detector is given by: 
\begin{eqnarray}
\widetilde{H} = \frac{1}{2} + \frac{1}{2} {\rm erf}  \left( \frac{ \widetilde{V} T - K }{ \sqrt{4 \widetilde{D} T} }   \right) 
\label{eq:gene:hit}
\end{eqnarray}
where ${\rm erf()}$ is the error function. We will now explain how we can make $\widetilde{H}$ to be approximately equal to $H$ in \eqref{eq:llr:h}. Note that the $\tanh$ function has been used to approximately compute ${\rm erf}$ in the past. 
Consider the worst case absolute error $e(\gamma) = \max_{u \in \mathbb{R} } |{\rm erf}(u) - \tanh(\gamma u) |$ for $\gamma \in \mathbb{R}$. The value of $\gamma$ that minimises $e(\gamma)$ is 1.198787 (which will be denoted by $\gamma_{\ast}$) and $e(\gamma_\ast) = 0.0197$  \cite{Chanson}. We can therefore approximately match $\widetilde{H}$ and $H$ by imposing that $\widetilde{H}(L_{0.5}) = H(L_{0.5})$ (= 0.5) and $\widetilde{H}(L) \approx H(L)$ for $L$ around $L_{0.5}$. The first requirement can be achieved by choosing $T$ to be:
\begin{eqnarray}
T & = & \frac{K}{\widetilde{V}(L_{0.5})}. \label{eq:match:half} 
\end{eqnarray}
We can meet the second requirement by choosing $\widetilde{D}(L_{0.5})$ so that:
\begin{eqnarray}
\gamma_\ast
\left. \frac{d \widetilde{V}}{dL} \right|_{L_{0.5}} 
\frac{1}{ \sqrt{\widetilde{V}(L_{0.5}) \widetilde{D}(L_{0.5}) } }  
= 
 \left. \frac{dV}{dL} \right|_{L_{0.5}} \frac{ \sqrt{K}  }{ D(L_{0.5}) } \label{eq:match:slope} 
\end{eqnarray}
See \smsec{\ref{app:tanh_erf}} for the derivation of these two requirements. 
Note that the above condition assumes that $D$ and $\widetilde{D}$ does not change much around $L_{0.5}$. This is consistent with the observations from our numerical study, see Figs.~\ref{fig:LLR_VD} and \ref{fig:p1_VD} in \smsec{\ref{app:fig}}. However, the condition \eqref{eq:match:slope} can be easily adjusted to take the variation of $D$ and $\widetilde{D}$ into consideration. Recall that there are $2n$ degrees-of-freedom in \ce{Y} and we have used $n$ of them so that the probability that all sites in \ce{Y} are bound is approximately equal to $Q$ in Sec.~\ref{sec:gene:mean}, therefore we can use the remaining degrees-of-freedom to enforce \eqref{eq:match:slope}. \\

An analytical expression to compute the mean first passage time $\widetilde{F}$ for $Z(t)$ to hit the boundary $K$ can also derived, see \smsec{\ref{app:promoter_fpt_upper}}. 

\subsection{Numerical results} 
\label{sec:num} 
This section presents some numerical results to illustrate the properties of the proposed method to emulate the hit rate of the SPRT detector. The SPRT is specified by 5 parameters: $g_+$, $g_-$, $L_0$, $L_1$ and $K$. We will keep the first four parameters the same and their values are given in \smsec{\ref{app:para}}. The values of $L_0$ and $L_1$ are taken from  \cite{Desponds:2020bla} which studies the possible use of SPRT in concentration detection of the morphogen Bicoid in \textsl{Drosophila} embryos. These values are based on the possible concentrations of Bicoid at the anterior-posterior boundary of \textsl{Drosophila} embryos. The binding rate $g_+$ for the binding site X is below the diffusion limited binding rate which is computed from estimated binding target dimension and diffusion coefficient taken from \cite{Desponds:2020bla}, see \smsec{\ref{app:para}}. We will use different values of $m$ which is the number of binding sites of the type X. We fix the total number of binding sites in the promoter \ce{X_{\rm m}-Y} to 6 and set the number of binding sites $n$ in the \ce{Y} part of the promoter as $n = 6 - m$. The quantity of 6 binding sites is also taken from \cite{Desponds:2020bla} as the hunchback gene in \textsl{Drosophila} is known to have at least 6 binding sites. However, in this paper, we use 6 binding sites because it is biologically realistic and we use it to set a resource limit on the promoter \ce{X_{\rm m}-Y}. We will use different values for the boundary $K$.  \\

\begin{figure}[t]
        \centering
        \includegraphics[scale = 0.45]{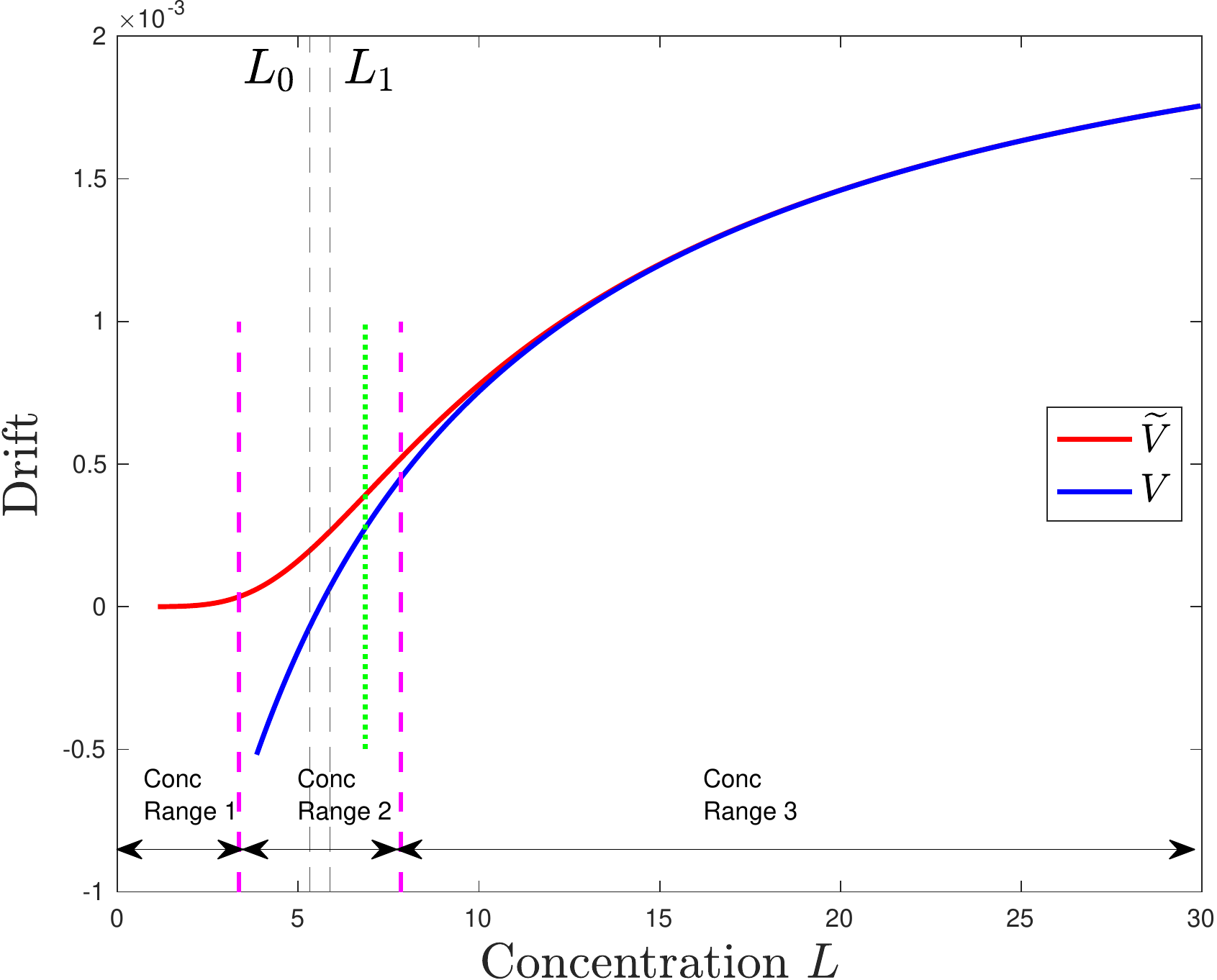} 
\caption{This plot compares the drift $\widetilde{V}$ of the mRNA production of Promoters 1 or 2, against that of the drift $V$ of the log-likelihood ratio \eqref{eq:llr:v}. 
}      
\label{fig:p1_drift}
\end{figure}

We first consider the case where $m = 1$ and hence $n = 5$. 
The \ce{Y} part of the promoter \ce{X_{\rm m}-Y} is specified by $2n$ parameters $a_i$ and $b_i$ for $i = 1, ..., n$. These parameters need to satisfy $\frac{a_i}{b_i} = \frac{1}{ L_{0.5} }$. We will construct two different \ce{Y} by choosing different values of $b_i$. The aim is to demonstrate that, for a given boundary $K$, we can use $\widetilde{D}(L_{0.5})$ to control the fit of $\widetilde{H}$ to $H$. In the first construction, which we will refer to as Promoter 1, we assume $b_i = g_-$ for all $i$; this means that $a_i = \frac{g_-}{L_{0.5}}$. For Promoter 2, we choose $b_i = 1.4 g_-$. Note that this construction implies that both Promoters 1 and 2 will have the same drift $\widetilde{V}$ but Promoter 2 will have a lower diffusivity $\widetilde{D}$ compared to Promoter 1 (see Fig.~\ref{fig:p1_VD} in \smsec{\ref{app:fig}}). Fig.~\ref{fig:p1_drift} compares the drift $\widetilde{V}$ of the promoters against the drift $V$ of the log-likelihood ratio. Note that the two drifts are equal for $L$ in Concentration Range 3 in Fig.~\ref{fig:p1_drift}. 
We remark that the \revisedversion{values} of $g_+$ and $g_-$ have been chosen such that the binding rates at all sites in Promoter 1 (i.e., including those sites in Y) are diffusion limited. However, for Promoter 2, one binding site does not have diffusion limited rate because a faster $b_i$ is used. \revisedversion{
We want to remark that the purpose of this example is to show that \eqref{eq:match:slope} can be used to select the boundary $K$ that can make $\widetilde{H}$ fit to $H$, so that the lack of diffusion limited binding rates for Promoter 2 does not affect the correctness of this study. }\\

\begin{figure}[t]
        \centering
        \includegraphics[scale = 0.45]{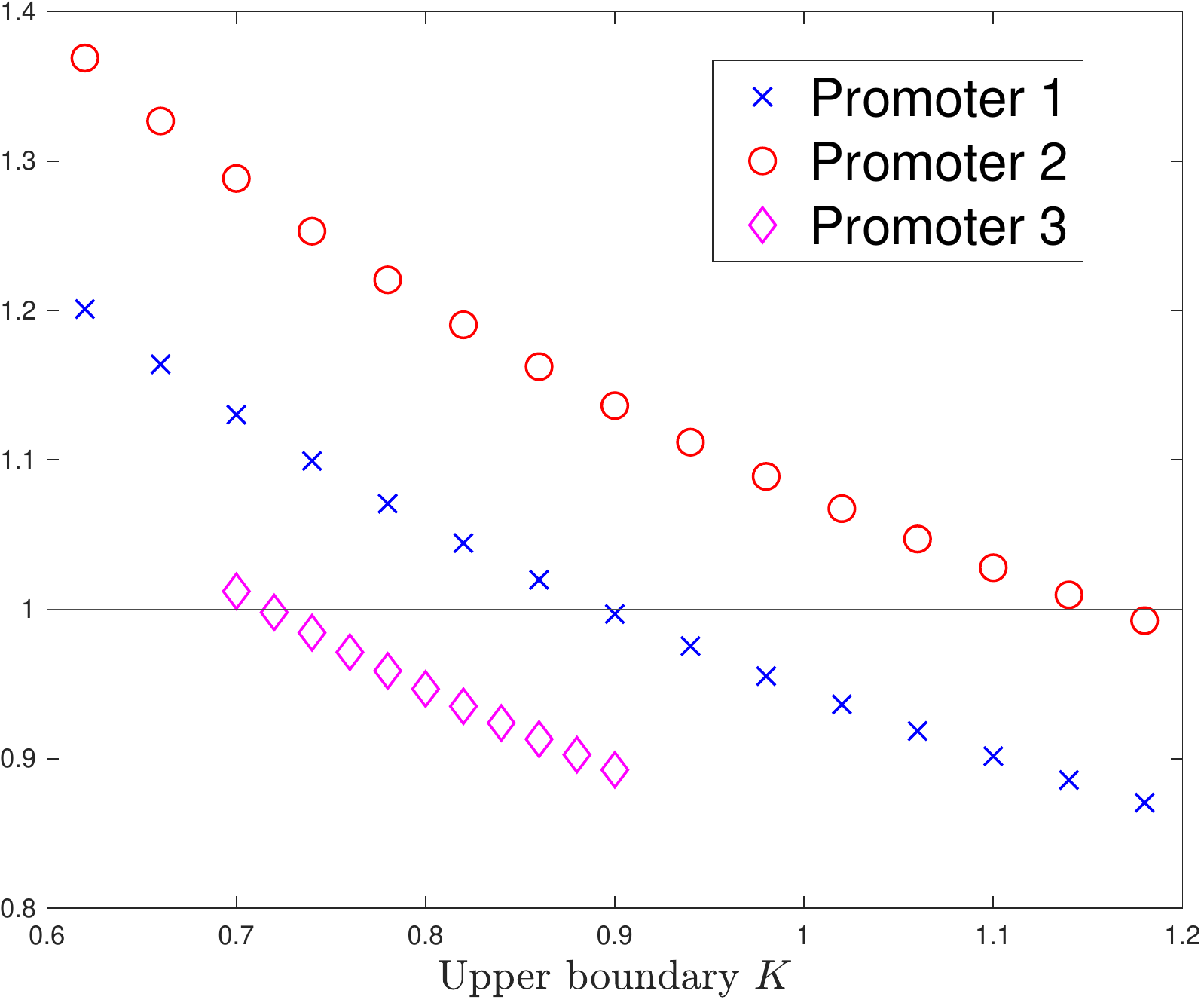} 
\caption{This figure compares the ratio of the left-hand side of \eqref{eq:match:slope} to its right-hand side for Promoters 1, 2 and 3.}      
\label{fig:p1_argerf}
\end{figure}

\begin{figure}[t]
        \centering
        \includegraphics[scale = 0.45]{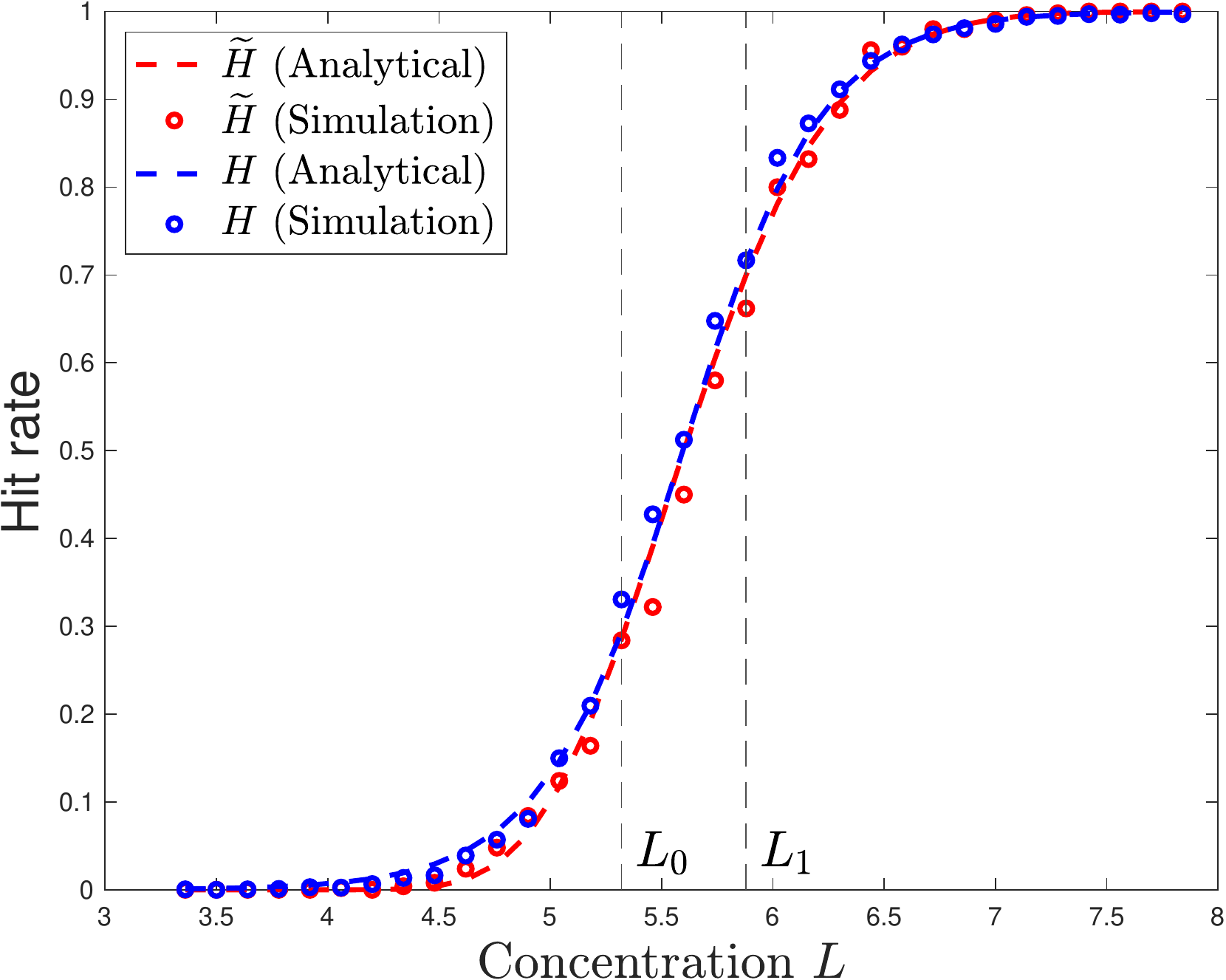} 
\caption{Comparing the hit rate $\widetilde{H}$ of Promoter 1 and that of SPRT ($H$). $K = 0.88$.}      
\label{fig:p1_hr}
\end{figure}

We first consider using Promoter 1 with the boundary $K = 0.88$. We calculate the duration $T$ of the transcription factor signal using \eqref{eq:match:half}. We will use the drift $\widetilde{V}$ and diffusivity $\widetilde{D}$ to check how well \eqref{eq:match:slope} holds. Fig.~\ref{fig:p1_argerf} plots the ratio of left-hand side of \eqref{eq:match:slope} to its right-hand side. It can be seen that the ratio is almost 1 for $K = 0.88$, so we should expect good fit of the hit rates $H$ and $\widetilde{H}$. Fig.~\ref{fig:p1_hr} compares the hit rate $H$ of the SPRT against the hit rate $\widetilde{H}$ of Promoter 1. It can be seen that the two hit rates are close to each other. Note that the simulation results, which are obtained from using Stochastic Simulation Algorithm (SSA) \cite{Gillespie:1977ww}, are close to those given by the analysis. (Simulation method is explained in \smsec{\ref{app:sim}}.) Fig.~\ref{fig:p1_argerf} shows that Promoter 1 should give poor fit for the hit rate for $K = 0.6$ and Fig.~\ref{fig:p1_hr_t4} (\smsec{\ref{app:fig}}) confirms that. The prediction of Fig.~\ref{fig:p1_argerf} is that Promoter 2 should be able to fit the hit rate for $K = 1.16$ and Fig.~\ref{fig:p2_hr_t4} (\smsec{\ref{app:fig}}) confirms that. This shows that it is possible to adjust the fit to the hit rate by adjusting the diffusivity $\widetilde{D}$. Note that the concentration range that we have focused on in Figs.~\ref{fig:p1_argerf}, \ref{fig:p1_hr_t4} and \ref{fig:p2_hr_t4} is Concentration Range 2 in Fig.~\ref{fig:p1_drift}. This is the concentration range that we need to focus on for fitting purposes, because the hit rate is either very close to 1 or 0 outside of this range. \\

\begin{figure}[t]
        \centering
        \includegraphics[scale = 0.45]{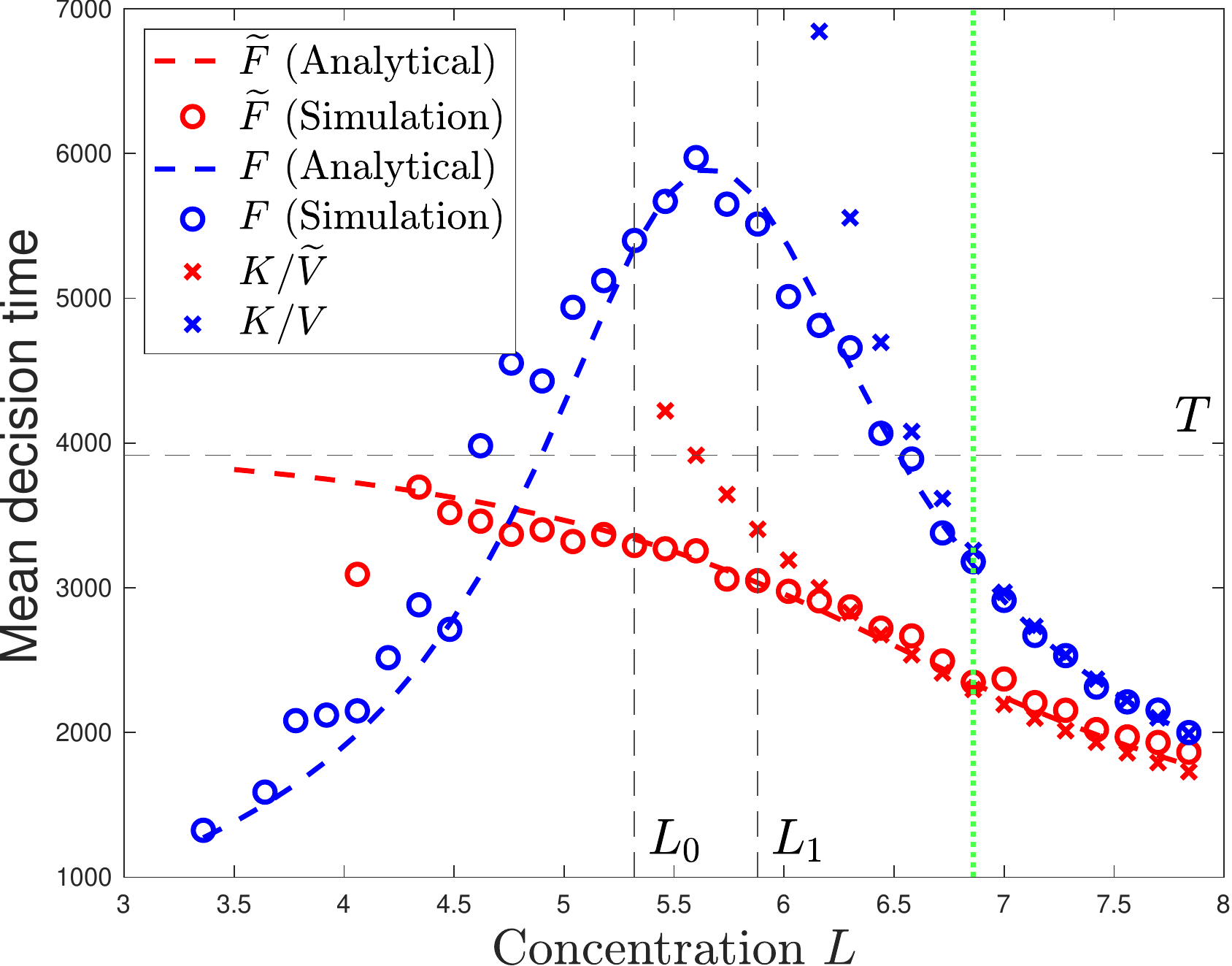} 
\caption{Comparing the mean decision time $\widetilde{F}$ of Promoter 1 and that of SPRT ($F$). $K = 0.88$.
}      
\label{fig:p1_fpt}
\end{figure}

We now return to Promoter 1 with $K = 0.88$ and consider its mean decision time or mean first passage time to hit the boundary $K$. Fig.~\ref{fig:p1_fpt} compares the mean decision time of the SPRT and that of Promoter 1. An interesting point to note is that for sufficiently high concentration levels in Fig.~\ref{fig:p1_fpt}, the mean decision time of Promoter 1 is \textsl{less} than that of the SPRT. Hence this is an advantage of the proposed transcription-based detector. Note that if $L$ is small, the hit rate will be low and if there is a hit, the mean decision time is almost equal to the signal duration $T$ as shown in Fig.~\ref{fig:p1_fpt}. The mean decision time for Promoter 2 with $K = 1.16$ has the same appearance, see Fig.~\ref{fig:p2_fpt} (\smsec{\ref{app:fig}}). We want to point out that these results do \textsl{not} contradict the optimality of the SPRT because our transcription-based detector is open-ended which implies that its mean time to decide for $L_0$ is infinity and is definitely poorer than that of the SPRT. \\

We see from Fig.~\ref{fig:p1_fpt} that when the concentration $L$ is near the high end, the mean decision times for SPRT and Promoter 1 are almost equal. In fact, this trend continues for higher concentration levels (such as Concentration Range 3 in Fig.~\ref{fig:p1_drift}), see Fig.~\ref{fig:p1_fpt_tail} (\smsec{\ref{app:fig}}). This is because for these concentration levels, $V \approx \widetilde{V}$ (Fig.~\ref{fig:p1_drift}), and $V \gg D$ and $\widetilde{V} \gg \widetilde{D}$ (Figs.~\ref{fig:LLR_VD}  and \ref{fig:p1_VD} in \smsec{\ref{app:fig}}), hence the mean decision time for both detectors tend to $\frac{K}{V}$. \\

In order to further explain why our transcription-based detector can achieve a lower mean decision time, we focus on those concentrations in Fig.~\ref{fig:p1_fpt} to the right of the green line. We can see that, for this concentration range, the mean decision times for the SPRT and our transcription-based detector are approximately equal to $\frac{K}{V}$ and $\frac{K}{\widetilde{V}}$, respectively. This means the drift is the main determining factor for the mean decision time. From Fig.~\ref{fig:p1_drift}, where the green vertical line corresponds to that in Fig.~\ref{fig:p1_fpt}, we can see that $\widetilde{V}$ is larger than $V$ in the concentration range that we are focusing on. Hence, the transcription-based detector achieves a lower mean decision time because it has a higher drift. The flip side of the above discussion is that, if the choice of the promoter binding and unbinding parameters results in $\widetilde{V}$ being smaller than $V$, then the mean decision time of the transcription-based detector will become higher than that of the SPRT.  \\

These numerical results explain why our transcription-based detector is able to emulate the hit rate of the SPRT detector. For high concentration $L$ (such as Concentration Range 3 in Fig.~\ref{fig:p1_drift}), the choice of binding and unbinding parameters in Sec.~\ref{sec:gene:mean} ensures the mean log-likelihood ratio ${\rm E}[L(t)]$ of the SPRT detector (which equals to $\widetilde{V}$) is approximately equal to the mean mRNA transcribed ${\rm E}[Z(t)]$ (which equals to $V$). Consequently, for this concentration range, we have $\widetilde{H} \approx H \approx 1$, as well as $\widetilde{F} \approx F \approx \frac{K}{V}$. 
For intermediate concentration $L$ (such as Concentration Range 2 in Fig.~\ref{fig:p1_drift}), 
we use \eqref{eq:match:half} and \eqref{eq:match:slope} to ensure that $\widetilde{H} \approx H$. We note that the approximation $\widetilde{H} \approx H$ is not a result of matching $V$ to $\widetilde{V}$ as Fig.~\ref{fig:p1_drift} shows that $V$ and $\widetilde{V}$ are different in Range 2.  In fact, $V$ is the drift of a log-likelihood ratio so its sign can be positive or negative in Range 2 depending on the value of concentration $L$, while $\widetilde{V}$ is non-negative because it is the production rate of mRNA. So, how could we have matched $H$ and $\widetilde{H}$ even when $V$ and $\widetilde{V}$ have such different numerical ranges? This is because $H$ in \eqref{eq:llr:h} and $\widetilde{H}$ in \eqref{eq:gene:hit} have similar mathematical forms, and the fact that the function ${\rm tanh}(\,)$ and ${\rm erf}(\,)$ have similar forms, so these similarities allow us to derive \eqref{eq:match:half} and \eqref{eq:match:slope} to match $\widetilde{H}$ to $H$. This discussion therefore reinforces a key idea of this paper which is the possibility of emulating the hit rate of the SPRT without having to compute the exact log-likelihood ratio. 
Finally for low concentration $L$ (such as Concentration Range 1 in Fig.~\ref{fig:p1_drift}), we have $\widetilde{H} \approx H \approx 0$. \\

\begin{figure}[!thbp]
        \centering
        \includegraphics[scale = 0.45]{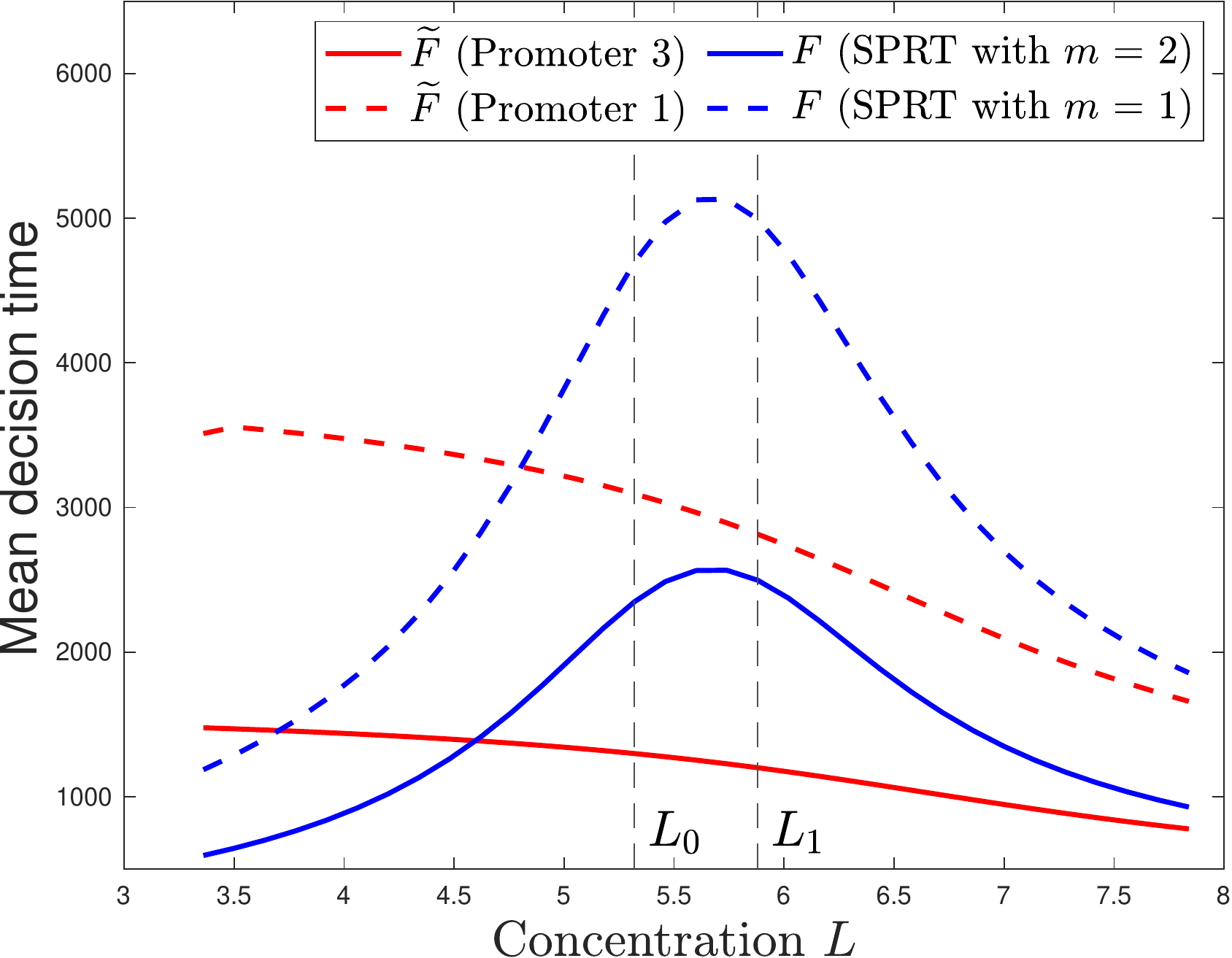} 
\caption{Comparing the mean decision time $\widetilde{F}$ of Promoters 1 and 3, as well as the mean decision time $F$ for SPRT  with $m = 1$ and $m = 2$. $K = 0.84$.
}      
\label{fig:p1_p3_fpt}
\end{figure}

We have so far focused on $m = 1$, we now consider $m \geq 1$. We know from Sec.~\ref{sec:sprt} that, if the same boundary $K$ is used for the SPRT detectors with $m = 1,2, ..$, then all these  detectors will have the same hit rate $H$ but their mean decision time will be inversely proportional to $m$. 
\revisedversion{
We want to see whether the transcription-based detector can imitate this type of behaviour. 
We continue to assume that there are a total of 6 binding sites, i.e., $m + n = 6$.
}
 We consider Promoter 3 with $m = 2$ and $n = 4$. We search for binding and unbinding rates $a_i$ and $b_i$ (for $i = 1, ..,4$) for the Y part of Promoter 3 to meet these criteria: (i) $\frac{a_i}{b_i} = \lambda$; (ii) The binding rates $a_i$ are diffusion limited; and (iii) The best boundary $K$ that Promoter 3 uses is almost the same as that of Promoter 1. (See \smsec{\ref{app:para}} for the parameter values for Promoter 3.) Fig.~\ref{fig:p1_argerf} shows the best $K$ for Promoter 3 is 0.7 but that for Promoter 1 is 0.88. We choose an in-between value of $K = 0.84$ and use it with both Promoters 1 and 3. (We find that values of $K$ between 0.82 and 0.86 give similar results.) Fig.~\ref{fig:p1_p3_hr} (\smsec{\ref{app:fig}}) shows that the hit rates for Promoters 1 and 3, and that of the corresponding SPRT with $K = 0.84$ are similar. Fig.~\ref{fig:p1_p3_fpt} shows that the mean decision time for Promoter 3 can be at least twice as fast as that of Promoter 1. \\

In \smsec{\ref{app:higher:m}}, we show that if the diffusion limited binding rate constraint is ignored, then the proposed transcription-based detector could continue to imitate the behaviour of SPRT for both $m = 3$ and $m = 4$ (i.e., $n = 3$ and $n = 2$ respectively)  but not for $m = 5$ (i.e., $n = 1$). This shows that it is possible to use as little as $n = 2$ binding sites to realise the threshold-hyperbolic function. In \smsec{\ref{app:higher:m}}, we explain why a smaller $n$ makes it difficult to meet diffusion limited binding rate constraint. However, note that the previous sentence assumes that $g_+$ and $g_-$ are given; if we conduct parameter search over $g_+$, $g_-$, $a_i$ and $b_i$, it is possible to find parameters which meet all the three criteria in the last paragraph for larger $m$.  

\section{Discussion and conclusions} 
\label{sec:discussion} 

\begin{figure}[!thbp]
        \centering
        \includegraphics[scale = 0.45]{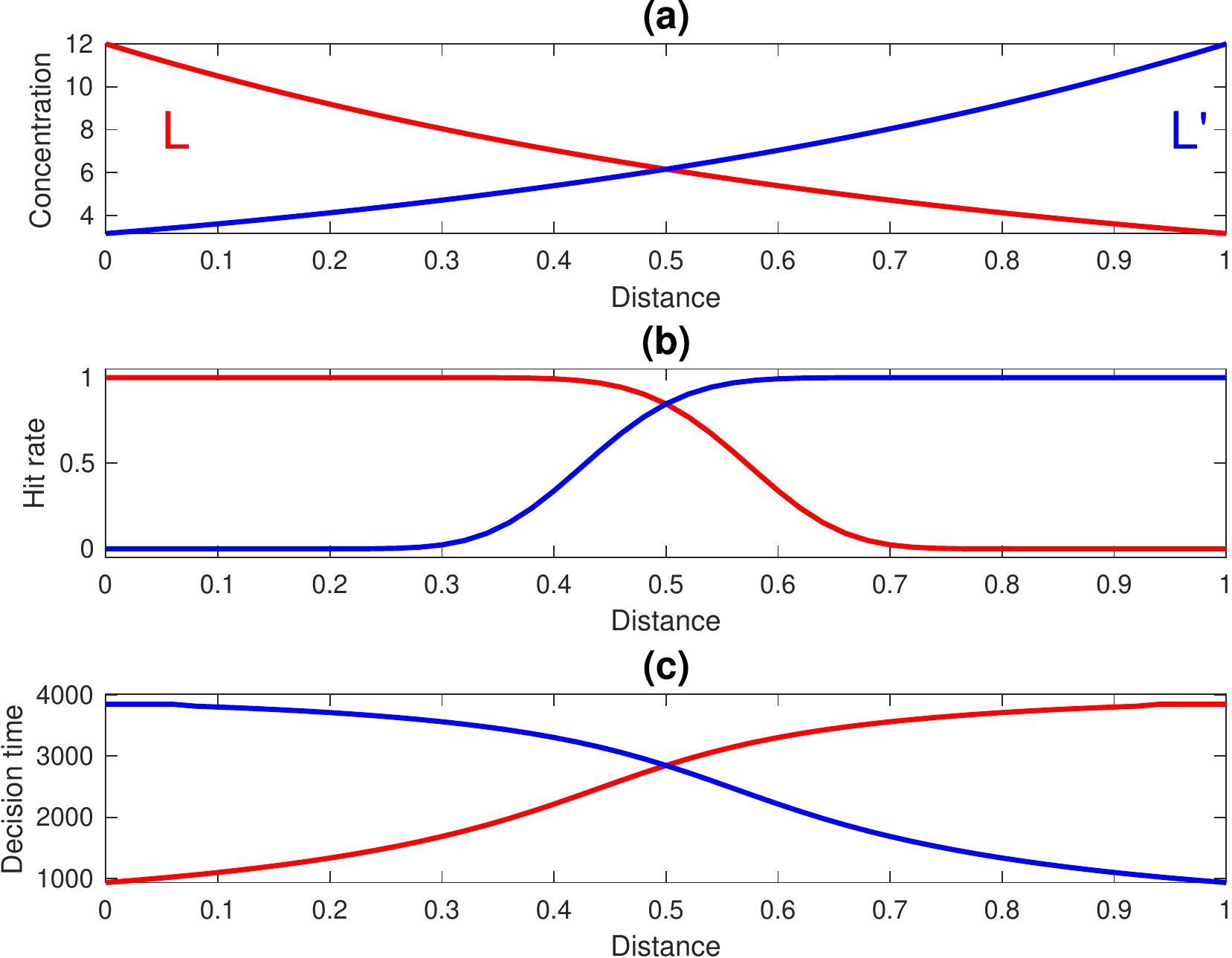} 
\caption{Using two transcription factors for decision making. The $x$-axis denotes distance along a line segment. (a) The concentration gradients for \ce{L} and ${\rm L}^{\prime}$. (b) Hit rates. (c) Mean decision time. 
}      
\label{fig:2tf}
\end{figure}

This paper shows that if we choose the binding and unbinding rates of a gene promoter appropriately, then we can use a finite duration transcription factor signal and the amount of mRNA transcribed to emulate the hit rate of a SPRT-based concentration detector. Furthermore, the mean response time of our transcription-based detector can be lower than or equal to that of the SPRT detector for a large concentration range. \\

A possible use of our result in synthetic biology is for cells to make decision according to an external concentration. We consider a 1-dimensional set up where cells are lined along a segment which we will denote by $[0,1]$. We assume that we can create two opposing concentration gradients of two transcription factors \ce{L} and ${\rm L}^{\prime}$ along the line segment, see Fig.~\ref{fig:2tf}(a). We assume that if the concentration of \ce{L} is above a target level, it can trigger a hit of our proposed transcription-based detector. We assume that this triggering will in turn cause the expression of another gene \ce{G}. Similarly, if the concentration of ${\rm L}^{\prime}$ is above a target level, it will cause a gene ${\rm G}^{\prime}$ to express. We further assume that the genes \ce{G} and ${\rm G}^{\prime}$ mutually repress each other so that only one of them will be expressed. Fig.~\ref{fig:2tf}(b) and (c) show respectively the hit rates and mean decision times of the transcription-based detectors; note these hit rates and decision times are hypothetical, and are based on Promoter 1 in Sec.~\ref{sec:num}. For the cells in the left half of the segment, the concentration of \ce{L} is higher and because the corresponding detector has a lower response time, this means \ce{G} will likely be expressed before ${\rm G}^{\prime}$ for this half of the segment; furthermore, since these genes mutually repress each other, the cells on the left half will likely express \ce{G}. Similarly, the cells in the right half will most likely express ${\rm G}^{\prime}$. Thus this hypothetical set-up will allow cells to make decisions according to an external concentration cue. \\

A key idea in this paper is to use a number of cooperative binding sites to approximately realise a threshold-hyperbolic function. We know from the linear framework for gene transcription in \cite{Ahsendorf:2014fc,Estrada:2016ct} that the behaviour of a set of cooperative binding sites can be modelled by a rational function of the concentration of the transcription factor. Therefore, from a mathematics point of view, one can view our work as using a rational function to approximate a 
mathematical function. This view point of rational function approximation can be used as a bridge to connect the computation carried out by living cells to their biochemical circuit realisation. 
We can use this view point to interpret the work in \cite{Olsman:2016cra} as using rational functions  to approximately realise logarithmic sensing in living cells. \\

We mention in Sec.~\ref{sec:intro} that a key distinction between \cite{Desponds:2020bla} and our work is that \cite{Desponds:2020bla} focuses on how the log-likelihood ratio in SPRT can be approximately computed while our work focuses on emulating the hit rate of SPRT. We now discuss a number of other differences. The work in \cite{Desponds:2020bla} considers a promoter consisting of 6 cooperative binding sites and assumes that transcription takes place when at least $k$ binding sites are bound where $k = 1, ..., 6$ is a parameter. It shows how SPRT can make use of the history of ON-OFF activity of the promoter for concentration detection. The work in this paper, which is based on independent and identical binding sites, can be considered as a special case of the promoter architecture considered in \cite{Desponds:2020bla}. It is possible to extend our work to the promoter architecture in \cite{Desponds:2020bla} by using the technique in \cite{Awan:2017fm} but the difficulty is to determine an approximate solution to a Bayesian filtering problem. We will leave this as future work. Another difference between \cite{Desponds:2020bla} and our work is that \cite{Desponds:2020bla} assumes that a common transcription rate is used among all the active promoter microstates while our work requires different transcription rates for different active microstate. However, we want to point out that our work can also make use of one common transcription rate, see \smsec{\ref{app:one_transcription_rate}} for the explanation. The key idea there is to make use of the bridge between rational function approximation and biochemical circuit realisation, which is discussed in the last paragraph and which we see is a useful connection. \\ 

In this paper, we assume that the concentration of the transcription factor is a constant over the time of detection. The same assumption is made in \cite{Siggia:2013dd,Desponds:2020bla}.  If we use the case of \textsl{Drosophila} embryo as a guide, this assumption means that the transcription factor needs to stay constant for a time-scale of 3 nuclear cycles \cite{Tran:2018jw}. In reality, the concentration of the transcription factor fluctuates over time. An interesting problem is to study the impact of this fluctuation on \revisedversion{both} the SPRT and our proposed transcription-based detector. This problem can be studied by assuming that the transcription factor is produced by a reaction-diffusion process and then couple this process with the binding-unbinding reactions in \eqref{cr:g:u0:all}. This combined reaction-diffusion process can be studied using the technique in \cite{Awan:2017fm} but the difficulty is to determine an approximate solution to a Bayesian filtering problem. We will leave this as future work.    
\\




\ifarxiv 
%

\else
	 \bibliography{nano,book,supp} 
\fi

\clearpage 
\pagebreak[4]
\newpage


   \renewcommand{\appendixname}{Supplemental Material}
\appendix
\renewcommand{\thesection}{\arabic{section}} 
\renewcommand{\thefigure}{S\arabic{figure}}
\setcounter{figure}{0}
\renewcommand{\thetable}{S~\arabic{table}}
\makeatletter
\setcounter{table}{0}
\renewcommand{\theequation}{S\thesection.\arabic{equation}}
\setcounter{page}{1}
\renewcommand{\thepage}{S\arabic{page}}

\renewcommand{\bibnumfmt}[1]{[S#1]}
    \renewcommand{\citenumfont}[1]{S#1}
    
\newcommand\SciteCoxMiller{S1}    
\newcommand\SciteChou{S2}          
\newcommand\SciteSiggia{S3}         
    \newcommand\SciteAhsendorf{S4}    
    \newcommand\SciteDesponds{S5}  

\hspace{2mm} \\
\section*{Supplemental Material to \textsl{Using transcription-based detectors to emulate the behaviour of sequential probability ratio-based  concentration detectors}} 

\section{Mean first passage time for SPRT to hit the upper boundary} 
\label{app:llr_fpt_upper}
The aim of this section is to derive \eqref{eq:llr_fpt}. We consider a Wiener process (or particle) with drift $V_W$ and diffusivity $D_W$. We assume that the particle is at $x_0$ at time 0. Consider a first passage time problem with absorbing boundaries at $-K$ and $K$ with $-K \leq x_0 \leq K$. Let $f_+(t | x_0)$ be the (un-normalised) probability density of the first hitting time of passing into $K$ and let $F_+(s | x_0) = \int_{t = 0}^\infty f_+(t | x_0) \exp(-st) dt$ be the Laplace transform of $f_+(s|x_0)$. According to [Chapter 5, \SciteCoxMiller], $F_+(s | x_0)$ is the solution of the differential equation
\begin{equation}
D_W \frac{d^2 F_+}{dx_0^2} + V_W \frac{dF_+}{dx_0} = s F_+
\end{equation}
with boundary conditions $F_+(-K) = 0$ and $F_+(K) = 1$. We can solve for $F_+(s | 0)$ using standard method. The mean passage time according to the un-normalised density is $-\left. \frac{d F_+}{ds} \right|_{s = 0}$, which can be worked out to be 
\begin{eqnarray}
 \frac{K_W}{V_W} \tanh \left( \frac{V_W K}{2 D_W} \right)  \; H 
\end{eqnarray}
where $H$ is the hit rate in \eqref{eq:llr:h}. The density $f_+$ is un-normalised because not all particles will hit the upper boundary. The normalisation constant is the hit rate $H$. 
We arrive at 
\begin{eqnarray}
\frac{K_W}{V_W} \tanh \left( \frac{V_W K}{2 D_W} \right)  
\end{eqnarray}
after normalisation. We now substitute $V_W = mV$ and $D_W = mD$ in the above expression, and we arrive at \eqref{eq:llr_fpt}. 

\section{Deriving \eqref{eq:llr:v}}
\label{app:proof:llr:v}
\begin{figure}[t]
        \centering
        \includegraphics[scale = 0.4]{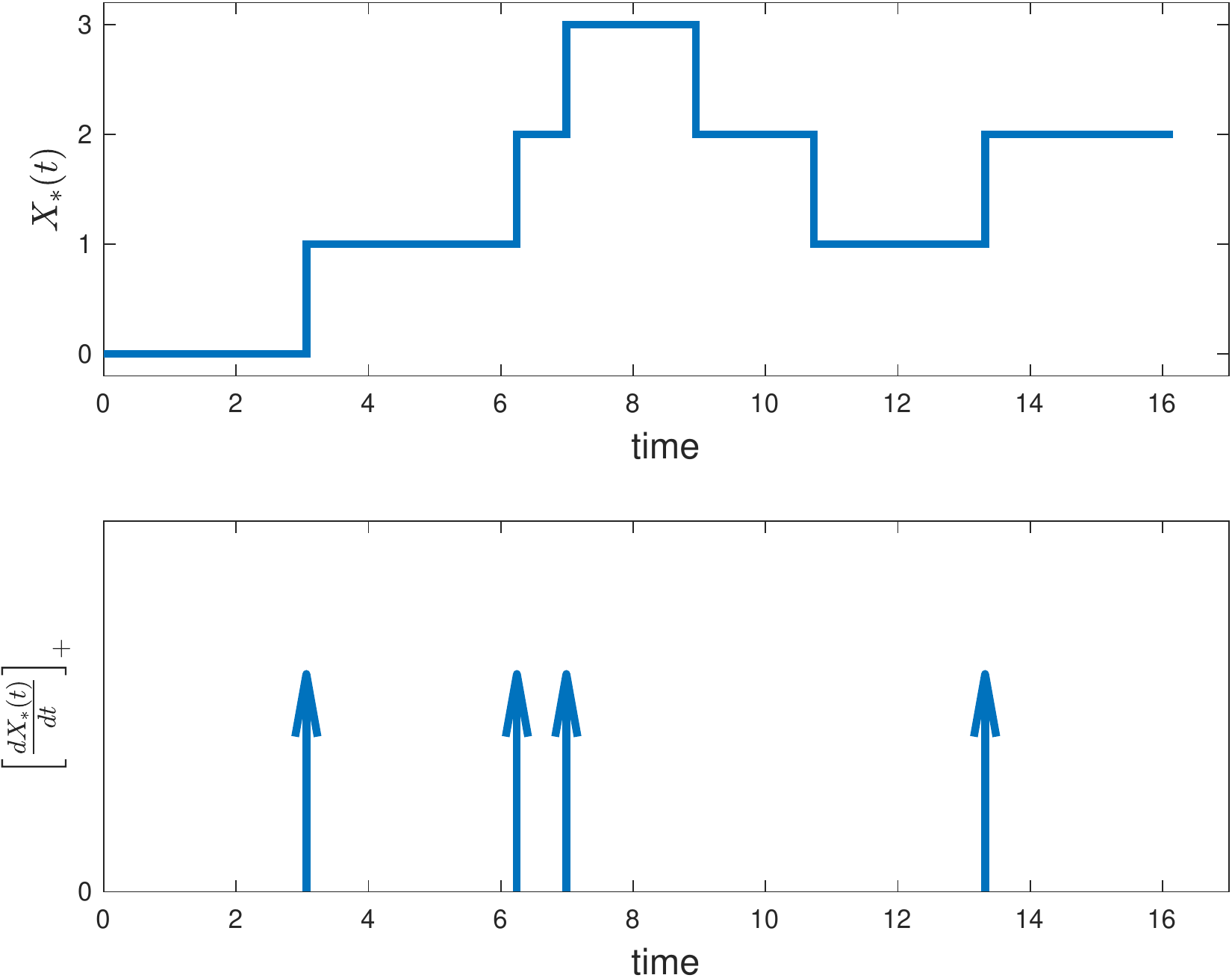} 
\caption{The top plot shows a sample realisation for $X_*(t)$. The bottom plot shows the $\left[ \frac{d X_*(t)}{dt} \right]_+$ of the $X_*(t)$ in the top plot. Note that each up arrow occurs at the time instant at which a binding occurs. }      
\label{fig:xstar_deri}
\end{figure}

In this section, we will derive \eqref{eq:llr:v} assuming that there are $m$ independent and identical binding sites of the type X. The measured datum $X_*(t)$ at time $t$ is the number of sites that are bound. Thus $X_*(t)$ is an integer in the interval $[0,m]$. The top plot in Fig.~\ref{fig:xstar_deri} shows a sample realisation of $X_*(t)$ for $m = 3$. \\

We show in [\SciteChou] that the computation of the log-likelihood ratio $R(t)$ in \eqref{eq:llr} requires us to determine the time instants at which a binding site switches from an unbound state to a bound state. We can see from the top plot in Fig.~\ref{fig:xstar_deri} that the binding time instants are those where there is a positive jump in the value of $X_*(t)$. Since $X_*(t)$ is piecewise constant, its time derivative $\frac{dX_*(t)}{dt}$ is a series of Dirac deltas where a positive Dirac delta corresponds to a binding instant.  Since we only require the positive Dirac deltas for the computation of the log-likelihood ratio, we use $\left[ \frac{d X_*(t)}{dt} \right]_+$ to retain those Dirac deltas that correspond to binding. The bottom plot in Fig.~\ref{fig:xstar_deri} shows the $\left[ \frac{d X_*(t)}{dt} \right]_+$ which corresponds to the $X_*(t)$ in the top plot in the figure. 
We show in [\SciteChou] that the evolution of the log-likelihood ratio $R(t)$ is given by: 

\begin{eqnarray}
\frac{dR(t)}{dt} & = & \left[ \frac{d X_*(t)}{dt} \right]_+ \log\left( \frac{L_1}{L_0} \right) - g_+ (m - X_*(t)) (L_1 - L_0) \nonumber \\ \label{app:eq:llr_ode}
\end{eqnarray}

In order to find the drift, we need to determine ${\rm E}[R(t)]$. By taking expectation on both sides of \eqref{app:eq:llr_ode}, we have:

\begin{eqnarray}
\frac{d{\rm E}[R(t)]}{dt} & = & {\rm E}\left[ \; \left[ \frac{d X_*(t)}{dt} \right]_+ \; \right] \log\left( \frac{L_1}{L_0} \right) - \nonumber \\ & &  g_+ (m - {\rm E}[X_*(t)]) (L_1 - L_0)  \label{app:eq:llr_ode_1}
\end{eqnarray}

The term ${\rm E}\left[ \; \left[ \frac{d X_*(t)}{dt} \right]_+ \; \right] $ can be interpreted as the mean binding rate and is given by $g_+ L (m - {\rm E}[X_*(t)])$. 
The term ${\rm E}[X_*(t)]$ is the mean number of sites that are bound and is equal to $\frac{m g_+ L}{g_- + g_+ L}$.  With these expressions, we have:

\begin{eqnarray}
\frac{d{\rm E}[R(t)]}{dt} & = & 
m 
\underbrace{
\frac{g_+ g_- }{g_- + g_+ L} \left(  L \; \log \left(\frac{L_1}{L_0} \right) - (L_1 - L_0)  \right) 
}_{= V {\mbox{ in \eqref{eq:llr:v}}}}
\nonumber \\ 
\label{app:eq:llr_ode_2}
\end{eqnarray}
which is the same as \eqref{eq:llr:v}. 

\section{Diffusivity of the log-likelihood ratio}
\label{app:llr:diff}
\revisedversion{The} aim of this section is derive a formula for the diffusivity $D$ of the log-likelihood ratio $R(t)$, which is defined as:
\begin{eqnarray}
D = \frac{1}{2} \frac{d {\rm Var}[R(t)]}{dt}
\end{eqnarray}
where ${\rm Var}[.]$ denotes the variance. 

\revisedversion{We first consider the case that there is only one binding site of the type X, i.e., $m$ = 1. An expression for the diffusivity $D$ for the case where $m = 1$ is given in [\SciteSiggia] and it is given by: 
} 
\begin{eqnarray}
D & = & \frac{1}{2} \frac{g_- g_+ L}{(g_- + g_+ L)^3} \times \nonumber \\
   &    & \bigg[  (g_+ (L_1 - L_0))^2 +  \nonumber \\
   &    & \frac{1}{2} \log^2\left( \frac{L_1}{L_0} \right) (g_-^2 + g_+^2 L^2) + \nonumber \\
   &    & g_+ (L_1 - L_0)  \log\left( \frac{L_1}{L_0} \right)  (g_- - g_+ L)  \bigg] 
   \label{app:eq:llr:D} 
\end{eqnarray}
For completeness, we provide an explanation on how the above expression on $D$ can be derived. \\

The ordinary differential equation \eqref{app:eq:llr_ode} shows the time evolution of the log-likelihood ratio $R(t)$ for a given trajectory of the state $X_*(t)$ of the binding site. By integrating \eqref{app:eq:llr_ode} assuming the initial condition $R(0) = 0$, we have 

\begin{eqnarray}
R(t) & = &  
\int_0^t \left[ \frac{d X_*(\tau)}{d\tau} \right]_+ d\tau 
\log\left( \frac{L_1}{L_0} \right) -
\nonumber \\ 
 & & g_+ (L_1 - L_0) \int_0^t (1 - X_*(\tau)) d\tau  \label{app:eq:llr_ode:int}
\end{eqnarray}

Let 
\begin{eqnarray}
 J_1(t) & = & \int_0^t \left[ \frac{d X_*(\tau)}{d\tau} \right]_+ d\tau  \\
 J_2(t) & = & \int_0^t  X_*(\tau) d\tau 
\end{eqnarray}
then we have:
\begin{eqnarray}
{\rm Var}[R(t)] & = & 
\log^2\left( \frac{L_1}{L_0} \right) {\rm Var}[ J_1(t) ]  + \nonumber \\
&  & (g_+ (L_1 - L_0))^2  {\rm Var}[ J_2(t) ] + \nonumber \\
&  & -2 \log\left( \frac{L_1}{L_0} \right) (g_+ (L_1 - L_0)) C_{J_1,J_2}(t,t) \nonumber \\
\label{app:eq:llr:D:varLLR}
\end{eqnarray}
where $C_{J_1,J_2}(t,s) = {\rm E}[J_1(t) J_2(s)] - {\rm E}[J_1(t)] {\rm E}[J_2(s)]$ is the correlation of $J_1(t)$ and $J_2(t)$. We remark that we do not need to consider the ``1" in the integral $\int_0^t (1 - X_*(\tau)) d\tau$ when we calculate the variance because it gives a deterministic drift. The next step is to work out the variances and correlation in the above equation. \\

We first consider ${\rm Var}[ J_1(t) ]$. In Eq.~(9) of the Supplementary Information in [\SciteSiggia], it was show that:
\begin{eqnarray}
{\rm Var}[ J_1(t) ]  & = & \frac{g_- g_+ L}{(g_- + g_+ L)^3} (g_-^2 + g_+^2 L^2) t 
\label{app:eq:llr:d:term1}
\end{eqnarray}
for large $t$. \\

We next consider ${\rm Var}[ J_2(t) ] $. It can be shown that 
\begin{eqnarray}
{\rm Var}[ J_2(t) ] = \int_0^t \int_0^t C_{X_*X_*}(s,u) ds du  
\label{app:eq:llr:var_J2} 
\end{eqnarray}
where $C_{X_*X_*}(s,u)$ is the cross-correlation of $X_*(s)$ and $X_*(u)$. It was shown in Eq.~(6) of the Supplementary Information in [\SciteSiggia] that:
\begin{eqnarray}
C_{X_*X_*}(s,u) = \frac{g_- g_+ L}{(g_- + g_+ L)^2} \exp(-(g_- + g_+ L) |s-u|). \nonumber 
\end{eqnarray}
Hence, after evaluating the double integral \eqref{app:eq:llr:var_J2}, we have:  
\begin{eqnarray}
{\rm Var}[ J_2(t) ] = \frac{2 g_- g_+ L}{(g_- + g_+ L)^3} t 
\label{app:eq:llr:d:term2}
\end{eqnarray}
for large $t$. \\

Finally we consider $C_{J_1,J_2}(t,s)$. It can be shown that:
\begin{eqnarray}
C_{J_1,J_2}(t,t) & = & \int_0^t C_{J_1 X_*}(t,s) ds 
\end{eqnarray}
where $C_{J_1 X_*}(t,s)$ is the cross-correlation between $J_1(t)$ and $X_*(t)$. We do not need to evaluate this integral because what we really need is the derivative of $C_{J_1,J_2}(t,t)$ with respect to $t$, which is equal to $C_{J_1 X_*}(t,t)$. We can obtain the expression of  $C_{J_1 X_*}(t,t)$ from Eq.~(7) of the Supplementary Information in [\SciteSiggia], so we have:
\begin{eqnarray}
\frac{d C_{J_1,J_2}(t,t)}{dt} & = & \frac{g_- g_+ L}{(g_- + g_+ L)^3} (g_- - g_+ L) 
\label{app:eq:llr:d:term3}
\end{eqnarray}

Next we use \eqref{app:eq:llr:d:term1}, \eqref{app:eq:llr:d:term2} and \eqref{app:eq:llr:d:term3} together with \eqref{app:eq:llr:D:varLLR} to obtain $\frac{1}{2} \frac{d {\rm Var}[R(t)]}{dt}$, which is the diffusivity. This exaplins how \eqref{app:eq:llr:D} can be derived.  \\

The expression \eqref{app:eq:llr:D} is the diffusivity when there is only one binding site of the type X. If there are $m$ identical binding sites of type X and the binding of the transcription factor L to these $m$ sites is independent, then the variance of the log-likelihood ratio when there are $m$ sites of X is simply $m$ times the variance of the log-likelihood ratio when $m = 1$. This follows from the fact that the variance of a sum of $m$ independent random variables is $m$ times the variance of each random variable. Thus, the diffusivity of the log-likelihood ratio for $m$ independent and identical binding sites of the type X is $mD$ where $D$ is given by \eqref{app:eq:llr:D}. 

\section{Deriving \eqref{eq:gene:v}}
\label{app:proof:gene:v}

The aim of section is to derive \eqref{eq:gene:v}. Note that the right-hand side of \eqref{eq:gene:v} is equal to the positive part of the right-hand side in \eqref{eq:llr:v}, which we will denote as $[m V]_+$. We will start from $[m V]_+$ and rewrite it into the form of the right-hand side of \eqref{eq:gene:v}.  

\begin{eqnarray}
& & [m V]_+ \nonumber \\
& = &m  \left[ \frac{g_+ g_- }{g_- + g_+ L} \left(  L \; \log \left(\frac{L_1}{L_0} \right) - (L_1 - L_0)  \right) \right]_\revisedversion{+} \nonumber \\
& = &m   \frac{g_+ g_- }{g_- + g_+ L} \left[ \left(  L \; \log \left(\frac{L_1}{L_0} \right) - (L_1 - L_0)  \right) \right]_\revisedversion{+} \nonumber \\
& = & g_- 
m \underbrace{\frac{g_+ L }{g_- + g_+ L}}_{= \bar{X}_*}
\left[
\log \left(\frac{L_1}{L_0} \right) - \frac{(L_1 - L_0)}{L}  
\right]_+
\nonumber \\
& = & g_- \; m \bar{X}_*  \; \log\left( \frac{L_1}{L_0} \right)
\left[
1 - \frac{\frac{L_1-L_0}{\log\left( \frac{L_1}{L_0} \right) }   }{L}
\right]_+ \nonumber \\
& = & g_- \; \log\left( \frac{L_1}{L_0} \right) \; m \bar{X}_*  \;
\left[
1 - \frac{L_{0.5}}{L}
\right]_+
\end{eqnarray} 
where the last expression is identical to the right-hand side of \eqref{eq:gene:v}.  \\

\section{Probability that all sites in \ce{Y} are bound}
\label{app:Yn} 
Assuming that $\frac{a_i}{b_i} = \lambda$ for $i = 1, ..., n$. Let $y_j$ be the probability that the state of the CTMC is \ce{Y_j}. The balance equations for the CTMC in \eqref{eq:g:y} are:
\begin{eqnarray}
\lambda L y_{j-1} = y_{j} \mbox{ for } j = 1, ..., n 
\end{eqnarray}
By solving these equations with $\sum_{j = 0}^n y_j = 1$, we have $y_n = f_n(L)$. 

\section{Calculating the drift $\widetilde{V}$ and diffusivity of $\widetilde{D}$} 
\label{app:promoter_VD}

The $m$ binding sites in \ce{X_{\rm m}} can be modelled by a continuous-time Markov chain (CTMC). Let $Q_{\rm X_{\rm m}}$ be the $(m+1)$-by-$(m+1)$ infinitesimal generator that describes this CTMC. Similarly, let $Q_{\rm Y}$ be the $(n+1)$-by-$(n+1)$ infinitesimal generator that describes \ce{Y}. Since the binding of \ce{L} to the binding sites in \ce{Y} is independent of those in \revisedversion{\ce{X_{\rm m}}}, we know from [\SciteAhsendorf] that the infinitesimal generator $Q_{\rm X_{\rm m}-Y}$ that describes the binding and unbinding behaviour of \ce{L} to \ce{X_{\rm m}-Y} is:
\begin{eqnarray}
Q_{\rm X_{\rm m}-Y} = Q_{\rm X_{\rm m}} \otimes I_{n+1} + I_{m+1} \otimes Q_{\rm Y} 
\end{eqnarray}
where $I_{n+1}$ is an identity matrix of size $(n+1)$-by-$(n+1)$ etc. and $\otimes$ is the Kronecker product. \\

Let $\pi$ be an $(m+1)(n+1)$ column vector which is the solution of the equation $Q_{\rm X_{\rm m}-Y} \pi = 0$. Therefore $\pi$ contains the steady state probabilities of the states of the CTMC that describes \ce{X_{\rm m}-Y}. Let $\rho$ be a $(m+1)(n+1)$ column vector which contains the transcription rates for the promoter states, i.e., the $i$-th element of $\rho$ contains the transcription rate of the promoter state that corresponds to $i$-th element of $\pi$. With these definitions, the drift $\widetilde{V}$ is given by $\rho^{\rm T} \pi$ where $^{\rm T}$ denotes matrix transpose. \\

It can be shown that the second moment of $Z(t)$ is given by:
\begin{eqnarray}
{\rm E}[Z(t)^2] = \int_0^t \int_0^t \rho^{\rm T} \exp(Q_{\rm X_{\rm m}-Y} |\tau_2 - \tau_1|) {\rm diag}(\pi) \; \rho \; d\tau_1 \; d\tau_2 \nonumber
\end{eqnarray}
where ${\rm diag}(\pi)$ turns the vector $\pi$ into a diagonal matrix. This integral can be evaluated by first computing the eigen-decomposition of $Q_{\rm X_{\rm m}-Y}$. After that, we will need to evaluate a number of integrals of the form
\begin{eqnarray}
\int_0^t \int_0^t \exp(-\xi |\tau_2 - \tau_1|) \; d\tau_1 \; d\tau_2
\end{eqnarray}
where $\xi$ is an eigenvalue of $Q_{\rm X_{\rm m}-Y}$. If $\xi \neq 0$, then this integral is approximately equal to $\frac{2 t}{\xi}$. Once ${\rm E}[Z(t)^2]$ has been computed, the diffusivity can be computed using standard methods.

\section{Deriving \eqref{eq:match:half} and \eqref{eq:match:slope}}
\label{app:tanh_erf} 
Given that ${\rm erf}(u) \approx \tanh(\gamma_\ast u)$ for $\gamma_\ast = 1.198787$, we can make $H$ and $\widetilde{H}$ approximately equal if
\begin{eqnarray}
\gamma_\ast \frac{ \widetilde{V} T - K }{ \sqrt{4 \widetilde{D} T } } = \frac{V K}{2 D} \label{app:eq:match} 
\end{eqnarray}
Recall that $V( L_{0.5} )  = 0$, hence we can use the choice of $T$ in \eqref{eq:match:half} to make $\widetilde{H}(L_{0.5}) = H(L_{0.5}) = 0.5$. This gives \eqref{eq:match:half}. \\

In the next step, we assume that both $D$ and $\widetilde{D}$ change slowly around $L = L_{0.5}$. This allows us to use the approximations $D(L) = D( L_{0.5} )$ and $\widetilde{D}(L) = \widetilde{D}( L_{0.5} )$ for $L$ close to $L_{0.5}$. Next, we differentiate both sides of \eqref{app:eq:match} with respect to $L$ and evaluate the derivative at $L = L_{0.5}$, and then use \eqref{eq:match:half} to eliminate $T$, we arrive at \eqref{eq:match:slope}. 


\section{Hit rate and mean first passage time for the amount of mRNA to first hit the upper boundary} 
\label{app:promoter_fpt_upper}

Let $Z_{\rm unrestricted}(t)$ be the trajectory of the amount of mRNA in the absence of the absorbing boundary $K$. This means $Z_{\rm unrestricted}(t) = Z(t)$ at any time $t$ where $Z(t)$ has not reached the boundary $K$. We now argue that we can determine the hit rate $\widetilde{H}$ of the transcription-based detector by using ${\rm Pr}[Z_{\rm unrestricted}(T) \geq K]$. In the absence of mRNA degradation, the trajectories for $Z_{\rm unrestricted}(t)$ are non-decreasing, so those trajectories that hit the boundary $K$ for some $t \leq T$ are exactly the same as those with $Z_{\rm unrestricted}(T) \geq K$. In addition, those trajectories $Z_{\rm unrestricted}(t)$ that reach the boundary $K$ are the same as those trajectories $Z(t)$ that hit the boundary $K$. Since $Z_{\rm unrestricted}(T)$ has a Gaussian distribution with mean $\widetilde{V} T$ and variance $2 \widetilde{D} T$, we have 
\begin{eqnarray}
\widetilde{H} = \int^{-\infty}_K p_{\rm N}(u ; \widetilde{V} T, 2 \widetilde{D} T) du
\end{eqnarray}
where $p_N(x ; \mu, \sigma^2)$ denotes the probability density of a Gaussian random variable $U$ with mean $\mu$ and variance $\sigma^2$. This integral can be rewritten in terms of the error function and we arrive at \eqref{eq:gene:hit}. 

Let $t$ be a time which is $\leq T$. It can be shown that the survival probability $S(t)$ for the absorption process is given by:
\begin{eqnarray}
S(t) = \int_{-\infty}^K p_{\rm N}(u ; \widetilde{V} t, 2 \widetilde{D} t) du. 
\end{eqnarray}
Given the survival probability $S(t)$, the mean first passage time is given by: 
\begin{eqnarray}
&     & - \int_{0}^T t \frac{d S(t)}{dt} dt \\
& = & \int_{0}^{T} S(t) - S(T) dt \label{eq:app:fpt_promoter}
\end{eqnarray}
Note that the above calculation is based on probability density which has not been normalised because not all trajectories will hit the upper boundary. The probability of hitting the upper boundary is the hit rate $\widetilde{H}$ or $1 - S(T)$. The mean first passage time for the promoter \ce{X_{\rm m}-Y} is therefore \eqref{eq:app:fpt_promoter} divided by this hit rate. This is the formula we use to analytically compute the mean first passage time of the promoter 

\section{Using microstates to realise the transcription rate}
\label{app:microstates}
Equation \eqref{eq:gene:v}, which is repeated below, shows that the mean cumulative amount of mRNA $M(t)$ transcribed by the transcription-based detector should obey: 
\begin{eqnarray}
\frac{dM(t)}{dt} =  
\underbrace{
g_- \; \log\left( \frac{L_1}{L_0} \right) 
}_{r}
\; 
m \; \bar{X}_*  \;
\underbrace{
\left[
1 - \frac{L_{0.5}}{L}
\right]_+
}_Q
\label{app:eq:mean:mRNA:m} 
\end{eqnarray}
where the right-hand side of the above equation should be interpreted as the mean transcription rate. We will explain how this mean transcription rate can be realised by using the thermodynamic theory of transcription based on microstates. \\
 
We will first discuss the case $m = 2$. In this case, the promoter has the form X$_{\rm 2}$--Y where the binding sites in Y are used to implement the threshold-hyperbolic function $Q$ as in Sec.~\ref{sec:gene:mean}. In order to realise the mean transcription rate in \eqref{eq:gene:v} or the right-hand side of \eqref{app:eq:mean:mRNA:m}, the promoter should behave as follows:
\begin{enumerate}
\item In the microstate where one of the X sites is bound and all sites in Y are bound, the mean transcription rate is $r$. The probability that the promoter is in these microstates is $2 \bar{X}_* (1- \bar{X}_*) Q$ recalling that $Q$ is the probability that all the binding sites in Y are bound. 
\item In the microstate where both of the X sites are bound and all sites in Y are bound, the mean transcription rate is $2r$. The probability that the promoter is in this microstats is $\bar{X}_*^2 Q$ 
\item No transcriptions occur in all other microstates. 
\end{enumerate}

Given the above microstate probabilities and mean transcription rates at these microstates, the mean transcription rate of the promoter X$_{\rm 2}$--Y is:
\begin{eqnarray}
2 \bar{X}_* (1- \bar{X}_*) Q \times r + \bar{X}_*^2 Q \times 2r & = & r \; 2 \bar{X}_* \; Q, 
\nonumber 
\end{eqnarray}
which is the desired mean transcription rate for $m = 2$. \\

For any integral $m$, the following transcription rules will be able to realise a mean transcription rate of $r \; m \bar{X}_* \; Q$. In this case, the promoter consists $m$ binding sites of type X and also a group of binding sites that constitutes Y. The transcription rules are: 

\begin{enumerate}
\item In the microstates where exactly $k$ sites of X are bound and all sites in Y are bound, we require a mean transcription rate of $k r$ where $k = 1, ..., m$. The probability that the promoter is in these microstates is $\binom{m}{k} \bar{X}_*^k (1- \bar{X}_*)^{m-k} Q$. 
\item No transcriptions occur in all other microstates. 
\end{enumerate}

It can be readily shown that
\begin{eqnarray}
\sum_{k = 1}^m \binom{\revisedversion{m}}{k} \bar{X}_*^k (1- \bar{X}_*)^{m-k} Q \; k r = r m \bar{X}_*  Q,  
\end{eqnarray}
which is the desired mean transcription rate when there are $m$ independent and identical sites of X.

\section{Parameter values}
\label{app:para} 
The parameter values for the numerical experiments are based on [\SciteDesponds] which studies the \textsl{hunchback} promoters in \textsl{Drosophila}. \\

The reference levels $L_0$ and $L_1$ are given respectively by 0.95 $\times$ $L_{\rm boundary}$ 
and 1.05 $\times$ $L_{\rm boundary}$ where $L_{\rm boundary}$ is the free concentration of the morphogen Bicoid at the anterior-posterior boundary of a \textsl{Drosophila} embryo. The value of $L_{\rm boundary}$ is 5.6 $\mu$m$^{-3}$. The values $L_0$, $L_1$ and $L_{\rm boundary}$ are taken from [\SciteDesponds]. As a result of these choices, we have $L_{\rm 0.5} = 5.5953$ and $\lambda = \frac{1}{L_{\rm 0.5}} = 0.1787$. Note that the value of $L_{\rm 0.5}$ is almost equal to that of $L_{\rm boundary}$ because $L_0$ and $L_1$ are close to each other. \\

The binding rate of the transcription factor is assumed to be diffusion limited. This upper bound $\mu_{\max}$ on the binding rate is calculated from $a D$ where $a$ and $D$ are respectively the binding target size and the diffusivity of the transcription factor. \revisedversion{
The values of $a$ and $D$ are taken from [\SciteDesponds] and their values are
$a = 3$nm and $D = 7.4$ $\mu$m$^2$s$^{-1}$. 
}
This gives a $\mu_{\max}$ of 0.0222$\mu$m$^3$s$^{-1}$. \\

For Promoter 1, we chose $g_+ = 0.0055$ ($\approx 0.25 \mu_{\max}$) and $g_- = 0.0248$ for binding site X. These values are used so that diffusion limited binding will hold for all 6 binding sites, as we will see in one moment. These choices result in a probability of 0.55 that the binding site X is bound when the transcription concentration is $L_{\rm boundary}$. For binding sites in Y, we chose $b_i = i g_-$ for $i = 1, ..., n$ where $n = 5$ is the number of binding sites in Y for Promoter 1. The values of $a_i$ are then determined from $\frac{a_i}{b_i} = \lambda$.  The requirement for diffusion limited binding means that $a_i$ must be no more than $(n - i + 1) \mu_{\max}$.  The numerical values of $\frac{a_i}{n-i+1}$ for $i = 1, 2, ..,5$ are 0.0009,  0.0022,  0.0044,  0.0089 and 0.0222, which are all no more than $\mu_{\max}$. \\

For Promoter 2, we used the same $g_+$ and $g_-$ for the binding site X. For the binding site Y, we chose $b_i = 1.4 \times i \times g_-$ and $a_i$ are then calculated from $\frac{a_i}{b_i} = \lambda$.  The numerical values of $\frac{a_i}{n-i+1}$ for $i = 1, 2, ..,5$ are 0.0012, 0.0031, 0.0062, 0.0124 and 0.0310. Note that one binding rate in Y exceeds the diffusion rate limit. Since our purpose of using Promoter 2 is to demonstrate that the relation in \eqref{eq:match:slope} holds, this is not an issue. \\

Promoter 3 consists of 2 independent binding sites X. Both of these binding sites have the same $g_+$ and $g_-$ as Promoter 1. We keep the total number of binding sites for Promoter 3 as 6, so the number of binding sites available for Y is 4. We chose binding rates $a_i = (n - i + 1) \mu_{\max}$ for $i$ = 1, 2, 3 and 4, i.e., all the binding rates to the binding sites in Y are at diffusion limit. The values of $b_i$ are computed by using $\frac{a_i}{b_i} = \lambda$. The resulting $\frac{b_i}{i}$ for $i$ = 1, 2, 3 and 4 are 0.4969,  0.1863, 0.0828 and 0.0311.



\section{Simulation} 
\label{app:sim}
The promoter \ce{X_{\rm m}-Y} can be modelled by a CTMC. Let \ce{X_{{\rm m},i}} be the state that $i$ out of $m$ type X sites are bound where $i = 0, ..., m$. The possible states of the CTMC are (\ce{X_{{\rm m},i}},\ce{Y_j}) for $i = 0,...,m$ and $j = 0,...,n$ where \ce{Y_j} is defined in Sec.\ref{sec:gene:mean}. In order to model the requirement that the transcription factor \ce{L} binds to all the \ce{X} sites and \ce{Y} independently, we impose the condition that the transition rates from (\ce{X_{{\rm m},i}},\ce{Y_j}) to (\ce{X_{{\rm m},i}},\ce{Y_{j+1}}) are the same for all $i$, as well as other similar constraints. \\

\revisedversion{
After a simulation run of the Stochastic Simulation Algorithm (SSA), we use the bounding state of the \ce{X}-part to obtain $X_*(t)$. We can then use this $X_*(t)$ to compute the log-likelihood ratio \eqref{eq:llr} by integrating \eqref{app:eq:llr_ode}.  
}
\\

Transcription of mRNA will occur in the states (\ce{X_{{\rm m},i}},\ce{Y_n}) for $i = 1,...,m$ at a mean rate of $i r$.


\section{Emulating SPRT for $m \geq 1$}
\label{app:higher:m}

In this section, we consider the emulation of the SPRT detectors for $m = 1, 2, ..., 5$. We assume all these detectors use a common value of boundary $K = 0.84$ which is the same as the one used for Promoters 1 and 3 in the main text. \\

For $m = 1$ and $m = 2$, we use respectively Promoter 1 and Promoter 3 in the main text. \\

For $m = 3, 4, 5$ (which correspond to $n = 3, 2, 1$ respectively), we search for binding and unbinding rates $a_i$ and $b_i$ so that the hit rate $\widetilde{H}$ of the transcription-based detectors best match the hit rate $H$ of the SPRT detector for $K = 0.84$. For $m = 3$ and $m = 4$, we are able to find $a_i$ and $b_i$ that can give good match provided that we do not impose the diffusion limited binding rate constraint; so this constraint is dropped for $m = 3$ and $m = 4$. However, for $m = 5$, which corresponds to $n = 1$, there does not appear to have enough degrees-of-freedom to realise a good match. \\

\begin{figure}[t]
        \centering
        \includegraphics[scale = 0.45]{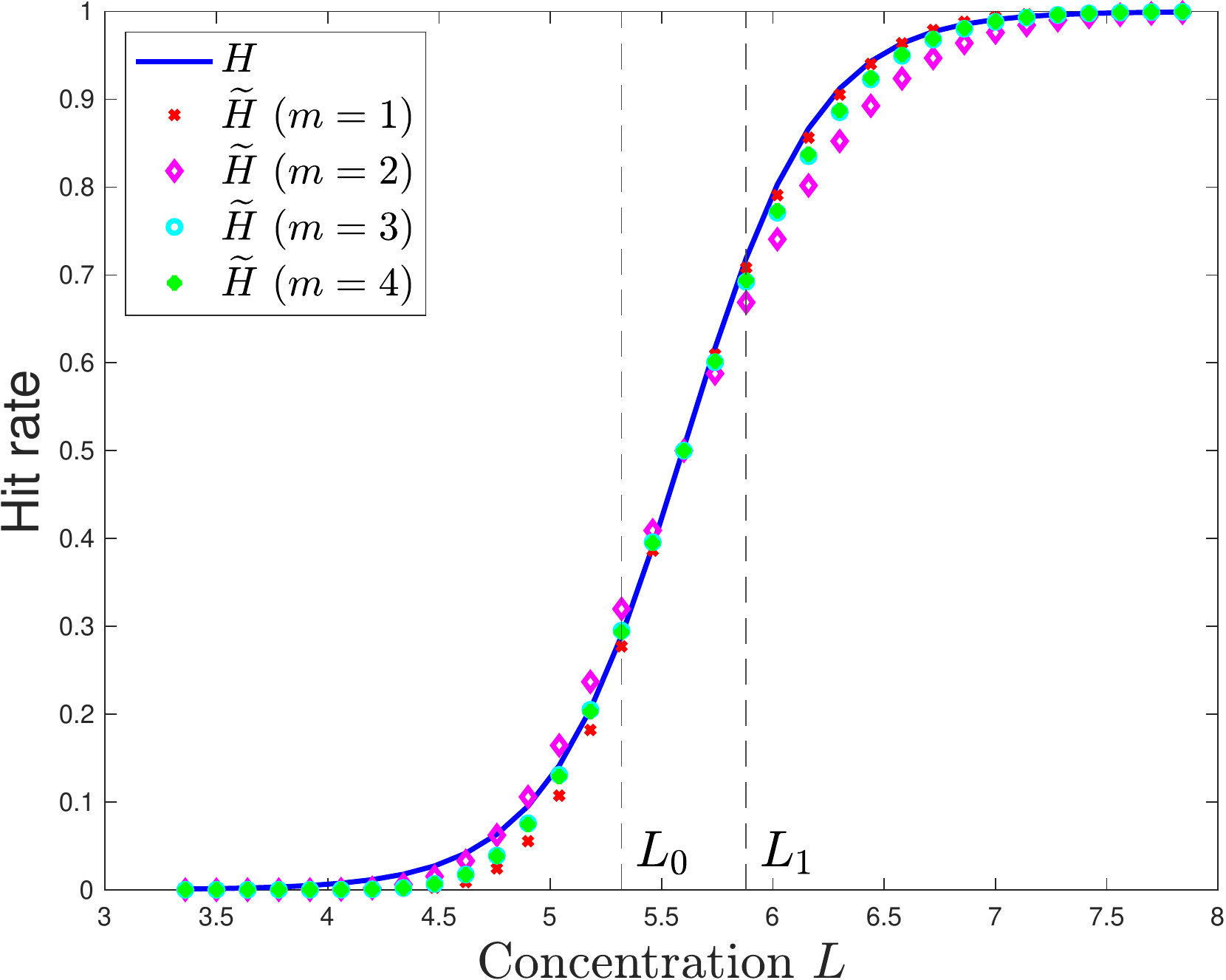} 
\caption{Comparing the hit rate $\widetilde{H}$ for $m = 1, 2, 3, 4$ and the hit rate $H$ of SPRT. $K = 0.84$.}      
\label{fig:m_bs_hr}
\end{figure}

\begin{figure}[t]
        \centering
        \includegraphics[scale = 0.45]{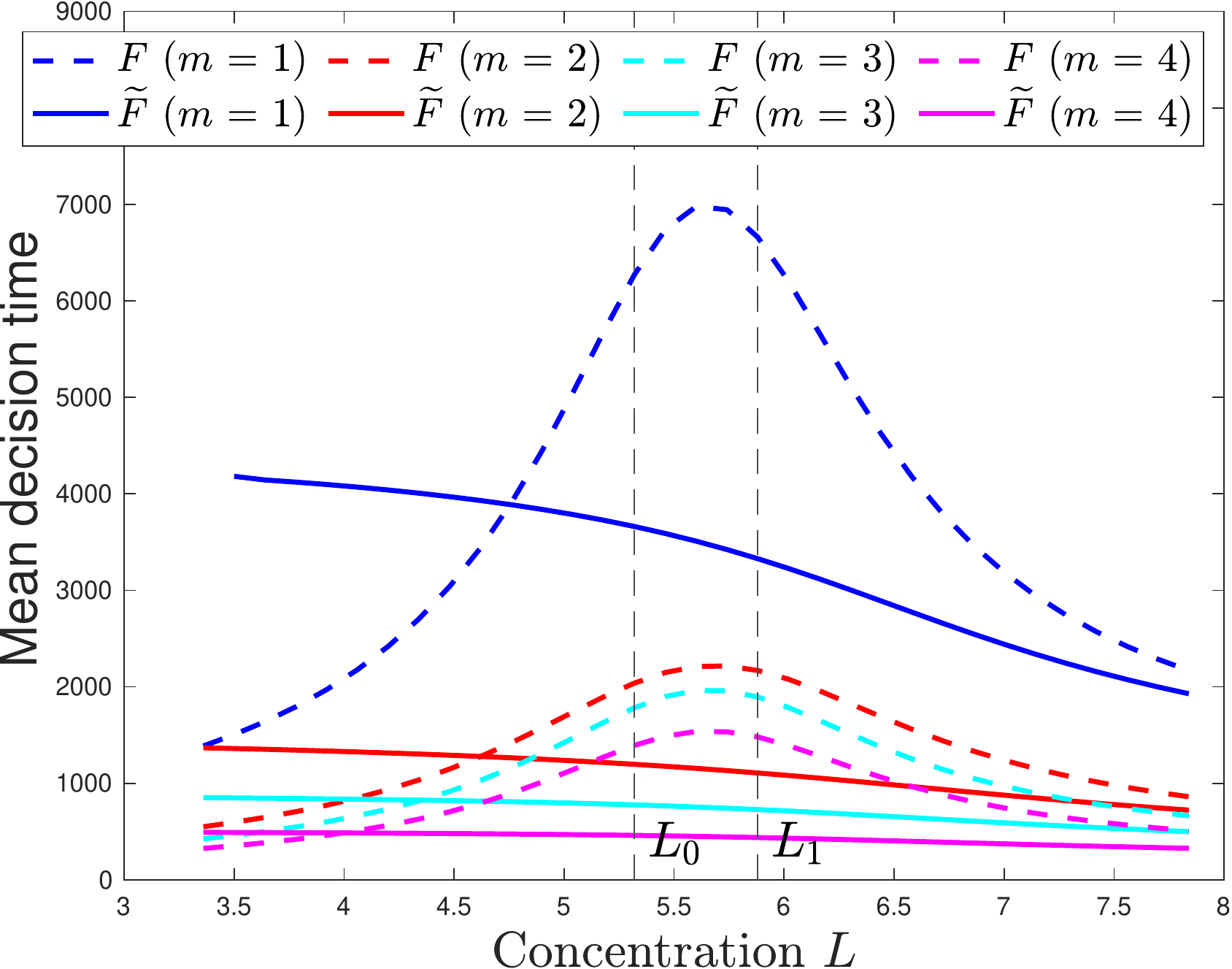} 
\caption{Comparing the mean decision time $\widetilde{F}$ for $m = 1, 2, 3, 4$ and the mean decision time $F$ of SPRT. $K = 0.84$.}      
\label{fig:m_bs_fpt}
\end{figure}

Fig.~\ref{fig:m_bs_hr} compares the hit rates $\widetilde{H}$ for the transcription-based detectors for $m = 1, ..., 4$ against that of the hit rate $H$ for the SPRT detector. We can see that the match is good. Fig.~\ref{fig:m_bs_fpt} compares the mean decision times of the $\widetilde{F}$ for the transcription-based detector for $m = 1, ..., 4$ against \revisedversion{the mean decision times} $F$ for the SPRT detectors for $m = 1, ..., 4$. We can see that higher value of $m$ leads to lower mean decision time. This show that two binding sites in the \ce{Y} part of the promoter \ce{X_{\rm m}-Y} are sufficient to approximate the threshold hyperbolic function if diffusion limited binding rate constrain is not taken into consideration. \\

We now explain why a faster binding rate is needed for a larger $m$ (or smaller $n$) to achieve a good match of the hit rate. Since $K$ is given, this means that the right-hand side of \eqref{eq:match:slope} is a constant. This implies that we require $\widetilde{D}(L_{0.5})$ to be proportional to 
\begin{eqnarray}
\left( 
\left. \frac{d \widetilde{V}}{dL} \right|_{L_{0.5}} \right)^2 
\frac{1}{ \widetilde{V}(L_{0.5})}
\end{eqnarray}
Recall that $\widetilde{V}(L) = r m X_{*}(L) f_n(L)$ so we will need to determine $f_n(L_{0.5})$. Since 
\revisedversion{
$\lambda = \frac{1}{L_{0.5}}$
}, we can use \eqref{eq:g:y} to show that $f_n(L_{0.5}) = \frac{1}{n+1}$. Further calculations lead to: 
\begin{eqnarray}
\widetilde{V}(L_{0.5}) & = & r m X_{*}(L_{0.5}) \frac{1}{n+1} \\
\left. \frac{d \widetilde{V}}{dL} \right|_{L_{0.5}}  & = & 
\widetilde{V}(L_{0.5}) \frac{1}{L_{0.5}} (1 - X_{*}(L_{0.5}) + \frac{n}{2} )
\end{eqnarray}
This means that $\widetilde{D}(L_{0.5})$ is required to scale with $n$ as follows:
\begin{eqnarray}
\widetilde{D}(L_{0.5}) & \propto & \frac{(6-n)(1 - X_{*}(L_{0.5}) + \frac{n}{2})^2}{n+1}
\end{eqnarray}
We can calculate the right-hand side of the above equation using our chosen parameters of $g_+$ and $g_-$ (which gives $X_{*}(L_{0.5}) = 0.55$), and for various values of $n$ which equals to $6 -m$. Our calculations show that we will need a larger diffusivity $\widetilde{D}(L_{0.5})$ for $m = 3, 4$ in comparison to $m = 1, 2$. Generally, one can get a higher diffusivity by using either a larger number of binding sites, or higher binding and unbinding rates. However, since we have already fixed the number of binding sites, so the only degree-of-freedom that we can use is to use a higher binding and unbinding rates. This explains why it is difficult to find binding rates within the limit to achieve \eqref{eq:match:slope} and hence the matching.


\section{Alternative realisation of \eqref{eq:gene:v}}
\label{app:one_transcription_rate}

A key idea in this paper is to use a rational function, which is realisable by a promoter, to approximate the threshold-hyperbolic function in \eqref{eq:gene:v}. This is so that we can approximately realise the mean transcription rate on the right-hand side of \eqref{eq:gene:v}. The aim of this section is to discuss an alternative method to realise this transcription rate. \\

Let us first point out that $\bar{X}_{*}$ and $Q$ are both functions of the transcription factor concentration $L$ and we have $0 \leq \bar{X}_{*} Q \leq 1$. The alternative method is to use a rational function to approximate the product $X_{*} Q$ in \eqref{eq:gene:v} instead of $Q$ alone. We know from [\SciteAhsendorf] that, for a promoter with multiple binding sites, the probabilities of the promoter microstates are rational functions in $L$ so we can try to fit a rational function to $X_{*} Q$. To make this discussion more concrete, we assume that we have decided on the number of binding sites in the promoter, and we will be optimising the binding and unbinding rates of this promoter to achieve a good fit. In order to perform this optimisation, we will collect all binding and unbinding rates into a parameter vector $\theta$. Let ${\cal S}$ be the set of all microstates of this promoter. Let $f_{\theta, s}$ be the probability that the promoter with parameter vector $\theta$ is in microstate $s \in {\cal S}$. We know that $f_{\theta, s}$ is a rational function in $L$ so our aim is to choose the $\theta$ and $s$ in order that $ f_{\theta, s} \approx X_{*} Q$. Let $s_*$ be the chosen microstate, then we required that the transcription rate be $m r$ in the microstate $s_*$. Note that instead of using one only microstate $s_*$, one may also choose a subset ${\cal S}_{\rm sub} \subset {\cal S}$ 
\revisedversion{of binding sites} 
to realise the approximation $\sum_{s \in {\cal S}_{\rm sub} } f_{\theta, s} \approx X_{*} Q$. In this case, we require that for all the microstates in ${\cal S}_{\rm sub}$, the transcription will take place at a rate of $mr$. \\

We want to finish this section with a remark. Note that in the main text, there is a clear separation in the roles for the \revisedversion{$(m+n)$}
binding sites. The $m$ binding sites, which provide the binding and unbinding history, are there to collect information on the concentration $L$ of the transcription factor. The $n$ binding sites are there to help to compute the threshold-hyperbolic function. However, if the method in the last paragraph is used, then this clear separation of roles will no longer apply. We will leave further study on this method as future work.

\section{Additional figures}
\label{app:fig}

\begin{figure}[h]
        \centering
        \includegraphics[scale = 0.45]{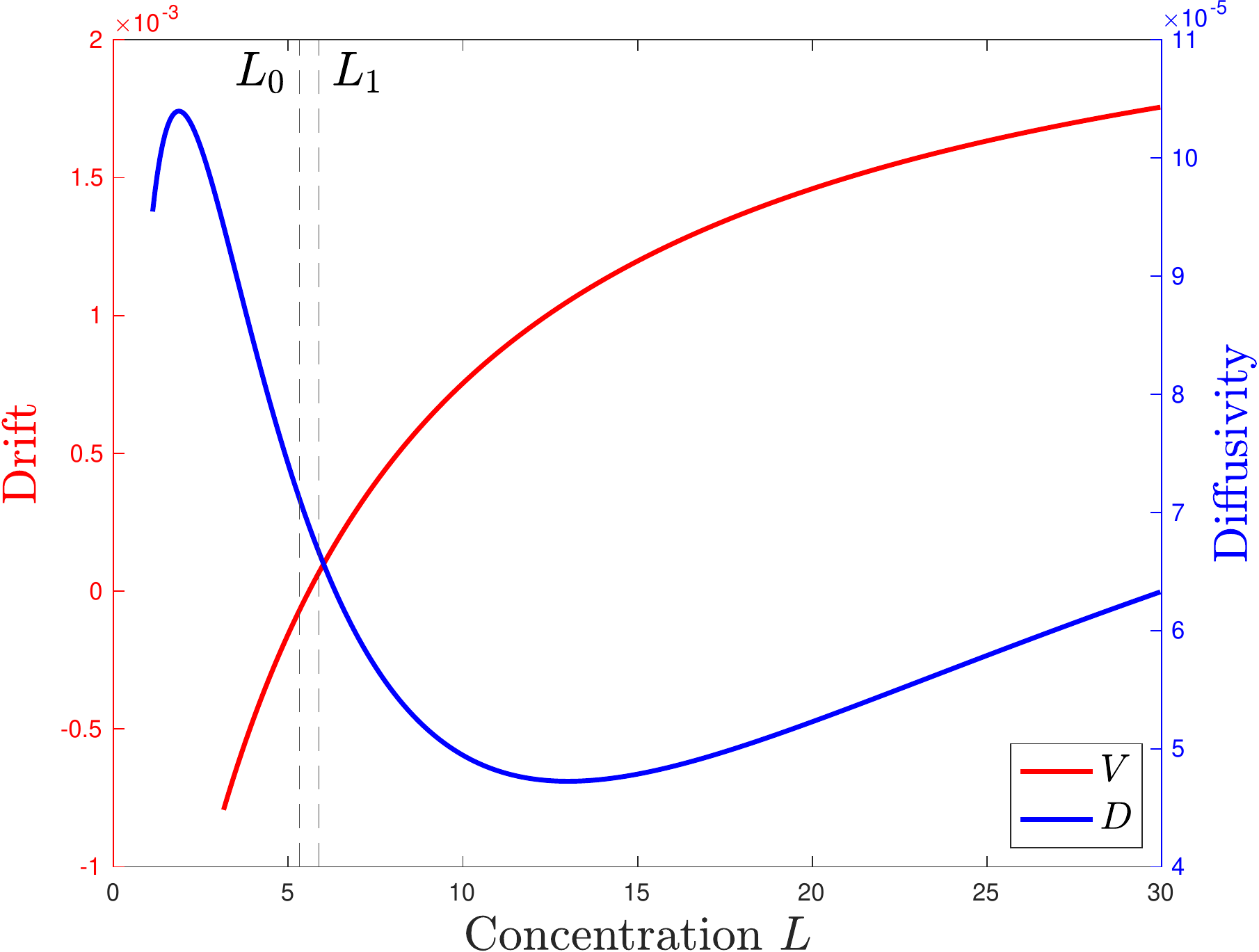} 
\caption{The drift $V$ and diffusivity $D$ of the SPRT. Note that the scales of $V$ and $D$ are different.}      
\label{fig:LLR_VD}
\end{figure}

\begin{figure}[h]
        \centering
        \includegraphics[scale = 0.45]{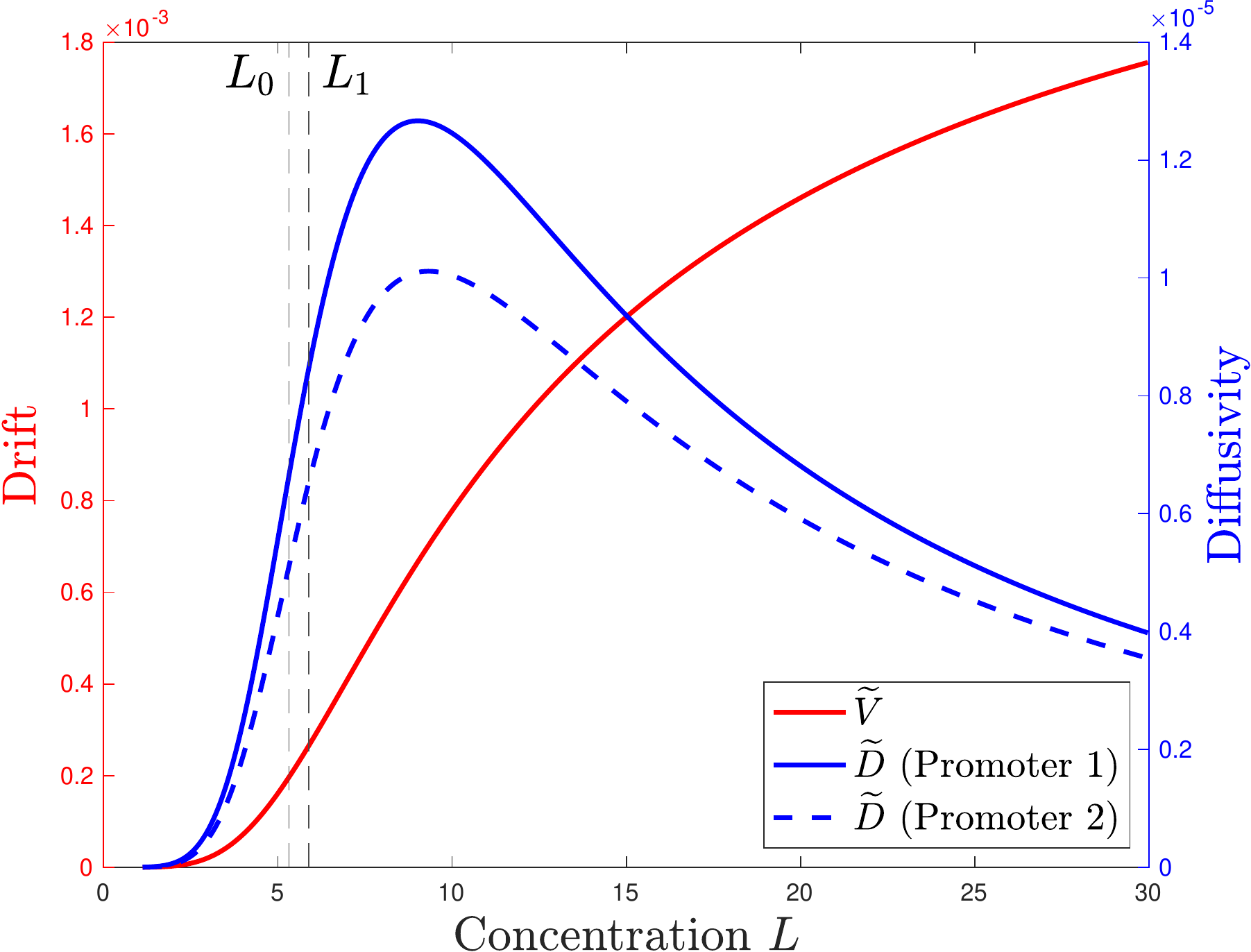} 
\caption{The drift $\widetilde{V}$ and diffusivity $\widetilde{D}$ of Promoters 1 and 2. Note that: (1) Both Promoters 1 and 2 have the same drift; (2) The scales of $\widetilde{V}$ and $\widetilde{D}$ are different.}      
\label{fig:p1_VD}
\end{figure}


\begin{figure}[h]
        \centering
        \includegraphics[scale = 0.45]{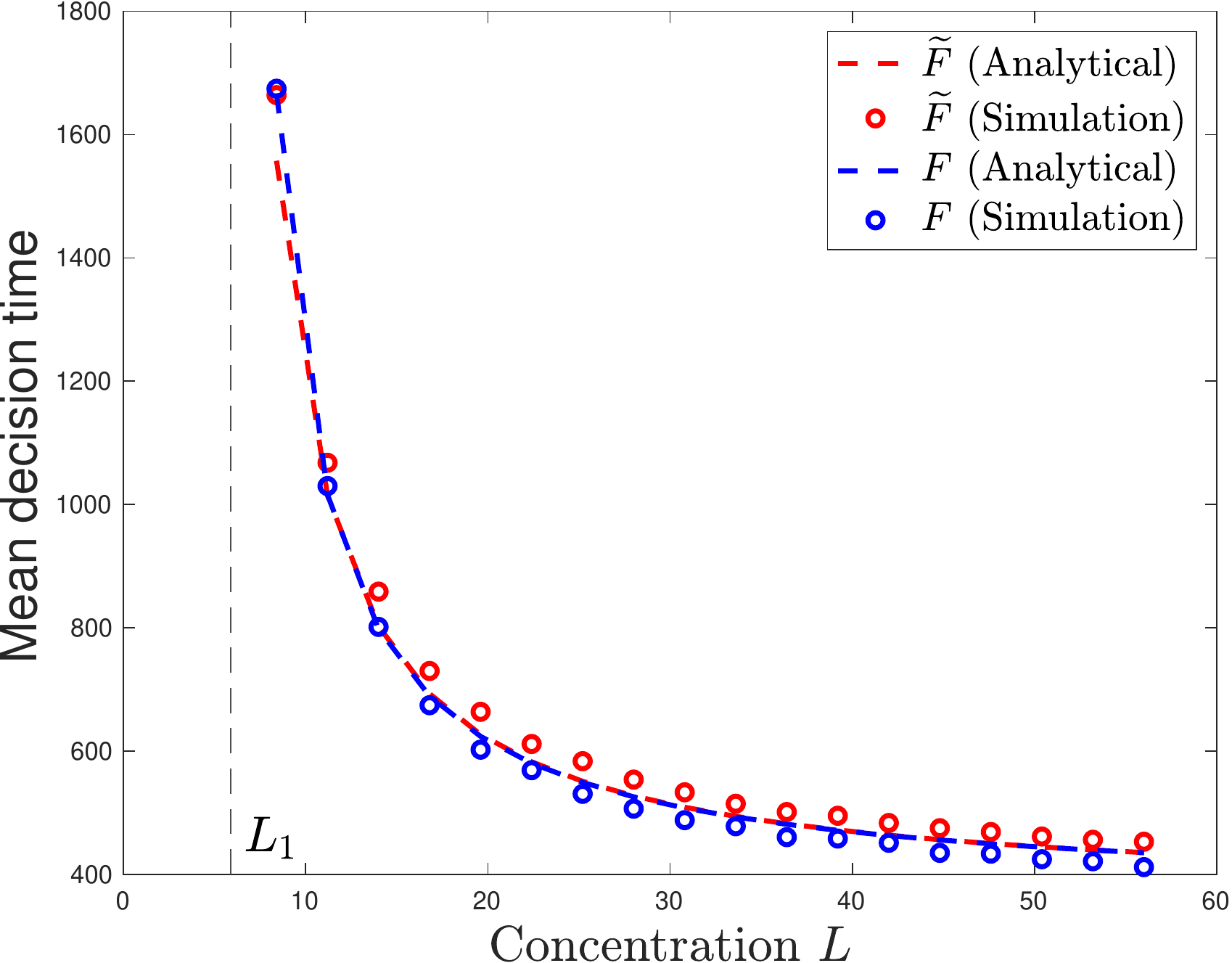} 
\caption{Mean decision time $\widetilde{F}$ for Promoter 1 for high concentration $L$. $F$ is the mean decision time for the SPRT. Boundary $K = 0.88$ .}      
\label{fig:p1_fpt_tail}
\end{figure}
	
\begin{figure}[h]
        \centering
        \includegraphics[scale = 0.45]{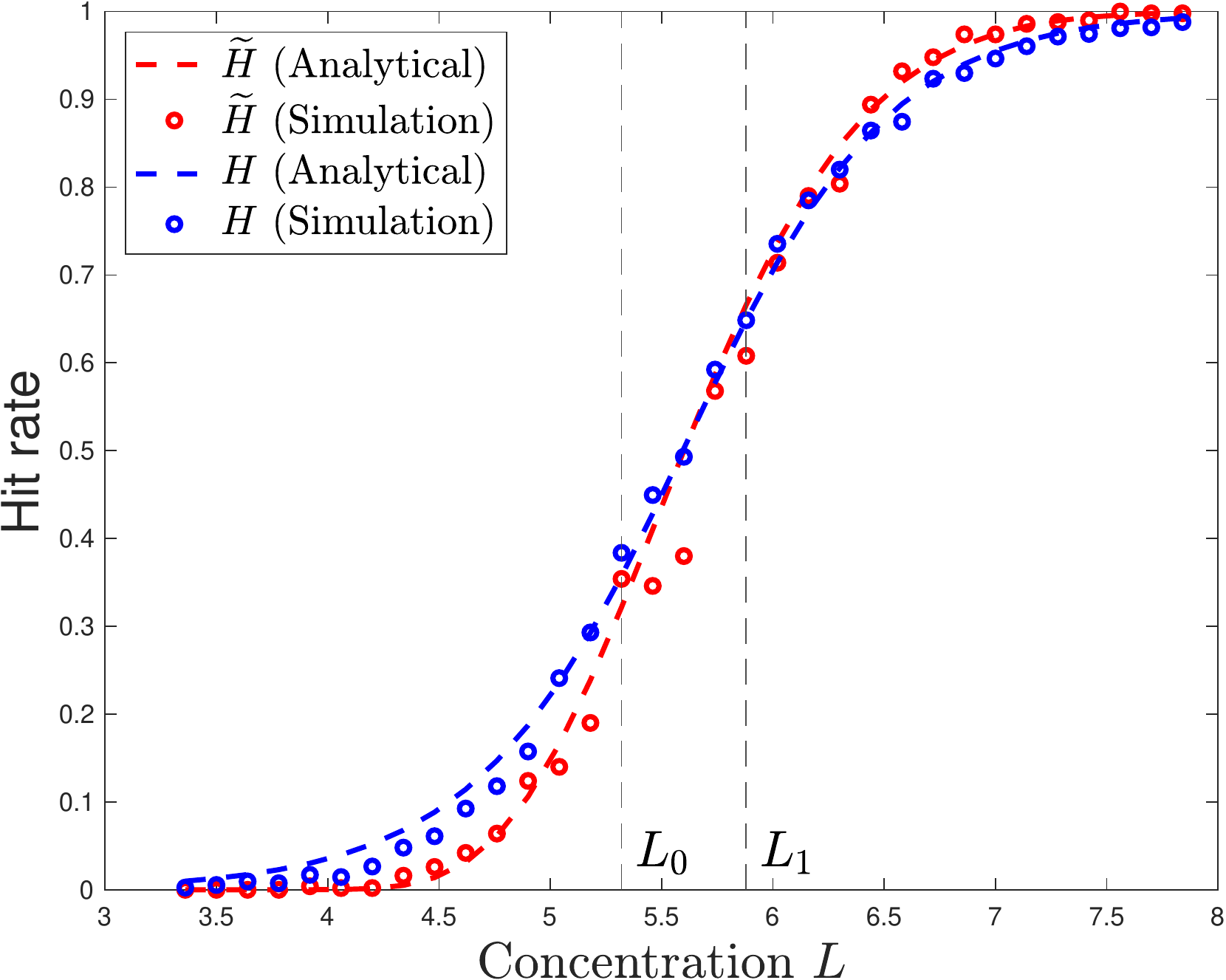} 
\caption{Hit rate $\widetilde{H}$ for Promoter 1 for boundary $K = 0.6$. $H$ is the hit rate for the SPRT.}      
\label{fig:p1_hr_t4}
\end{figure}

\begin{figure}[h]
        \centering
        \includegraphics[scale = 0.45]{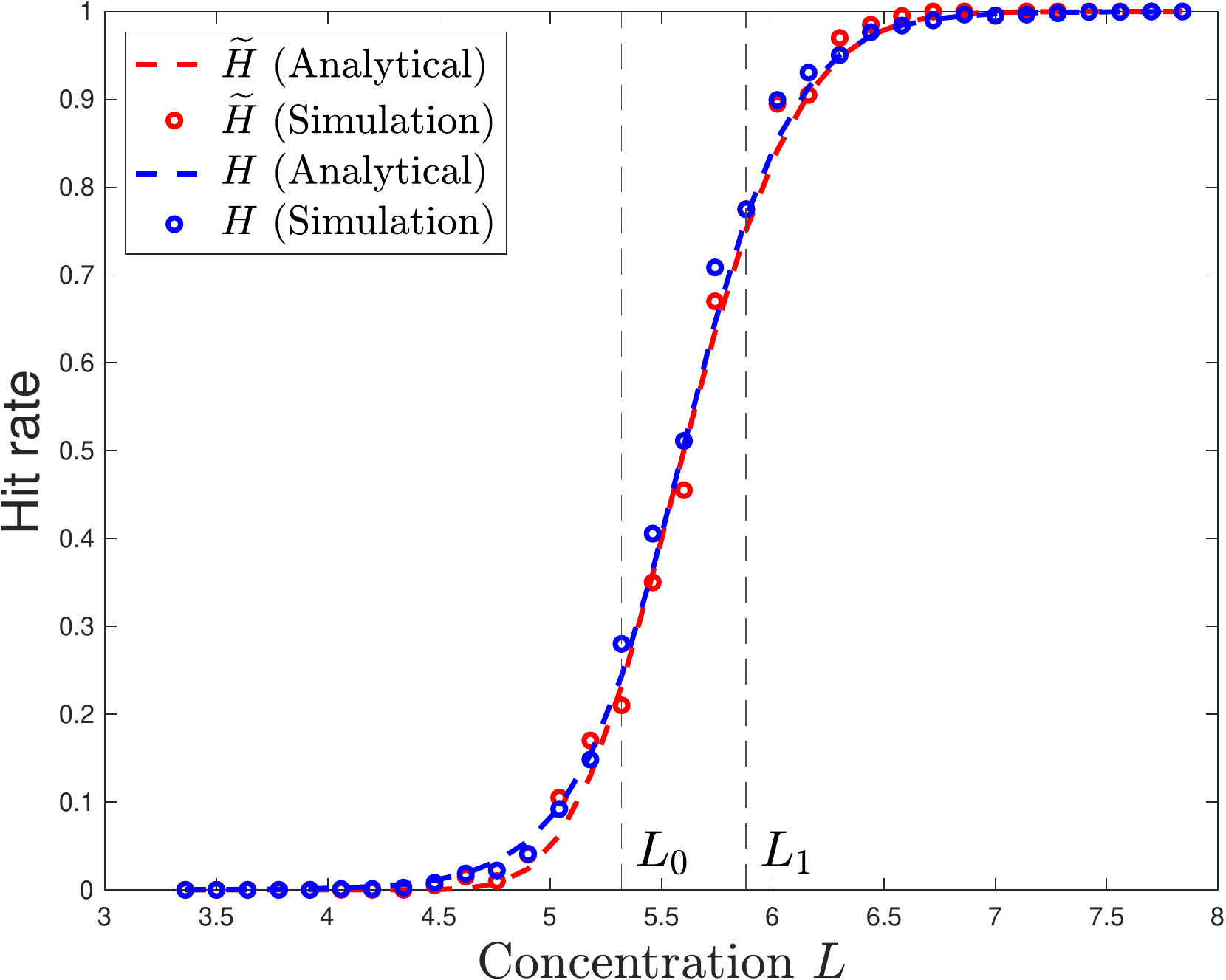} 
\caption{Hit rate $\widetilde{H}$ for Promoter 2 for boundary $K = 1.16$. $H$ is the hit rate for the SPRT.}      
\label{fig:p2_hr_t4}
\end{figure}

\begin{figure}[h]
        \centering
        \includegraphics[scale = 0.45]{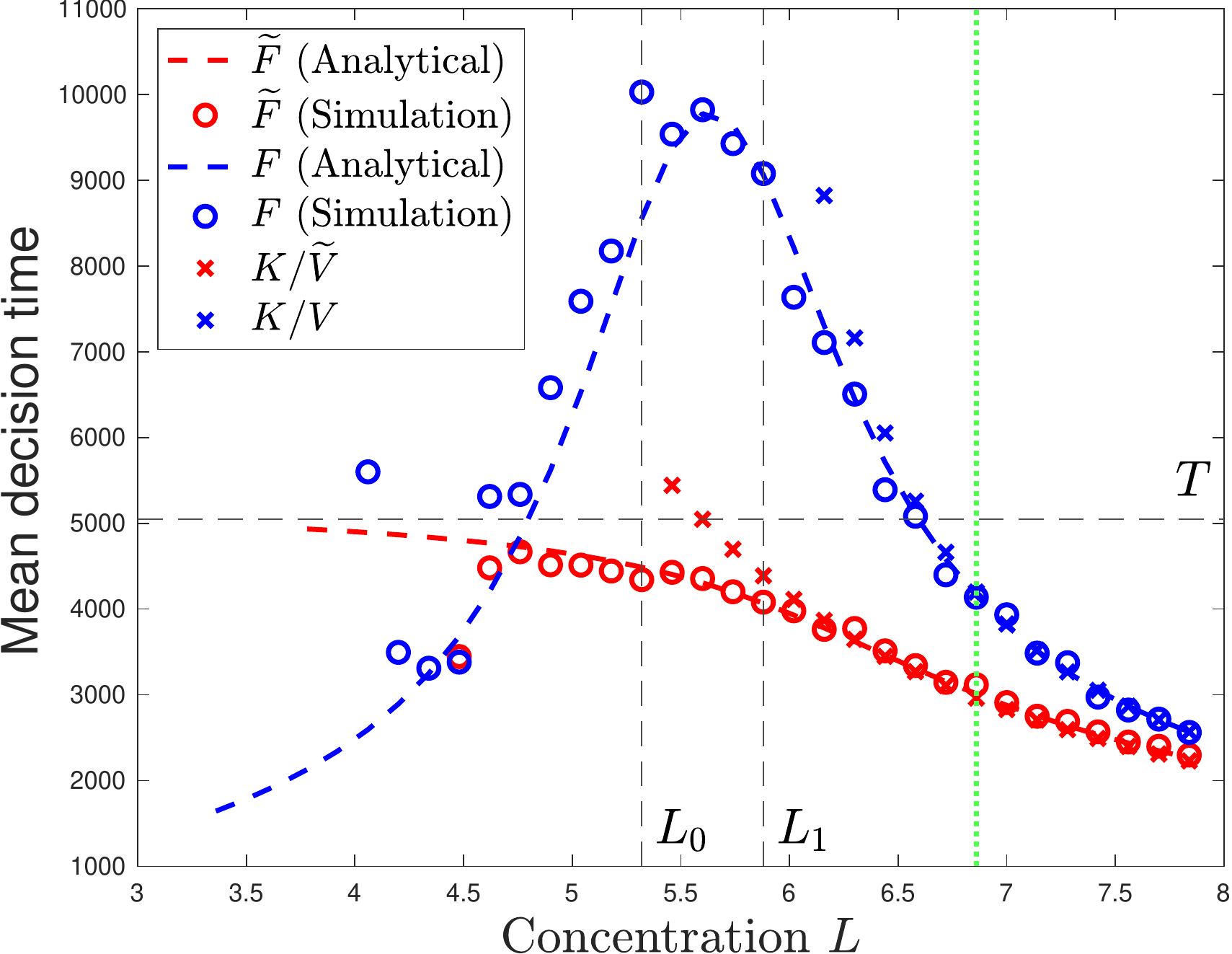} 
\caption{Mean decision time $\widetilde{F}$ for Promoter 2 for concentration around $L_{0.5}$. $F$ is the mean decision time for the SPRT.  Boundary $K = 1.16$ .}      
\label{fig:p2_fpt}
\end{figure}

\begin{figure}[t]
        \centering
        \includegraphics[scale = 0.45]{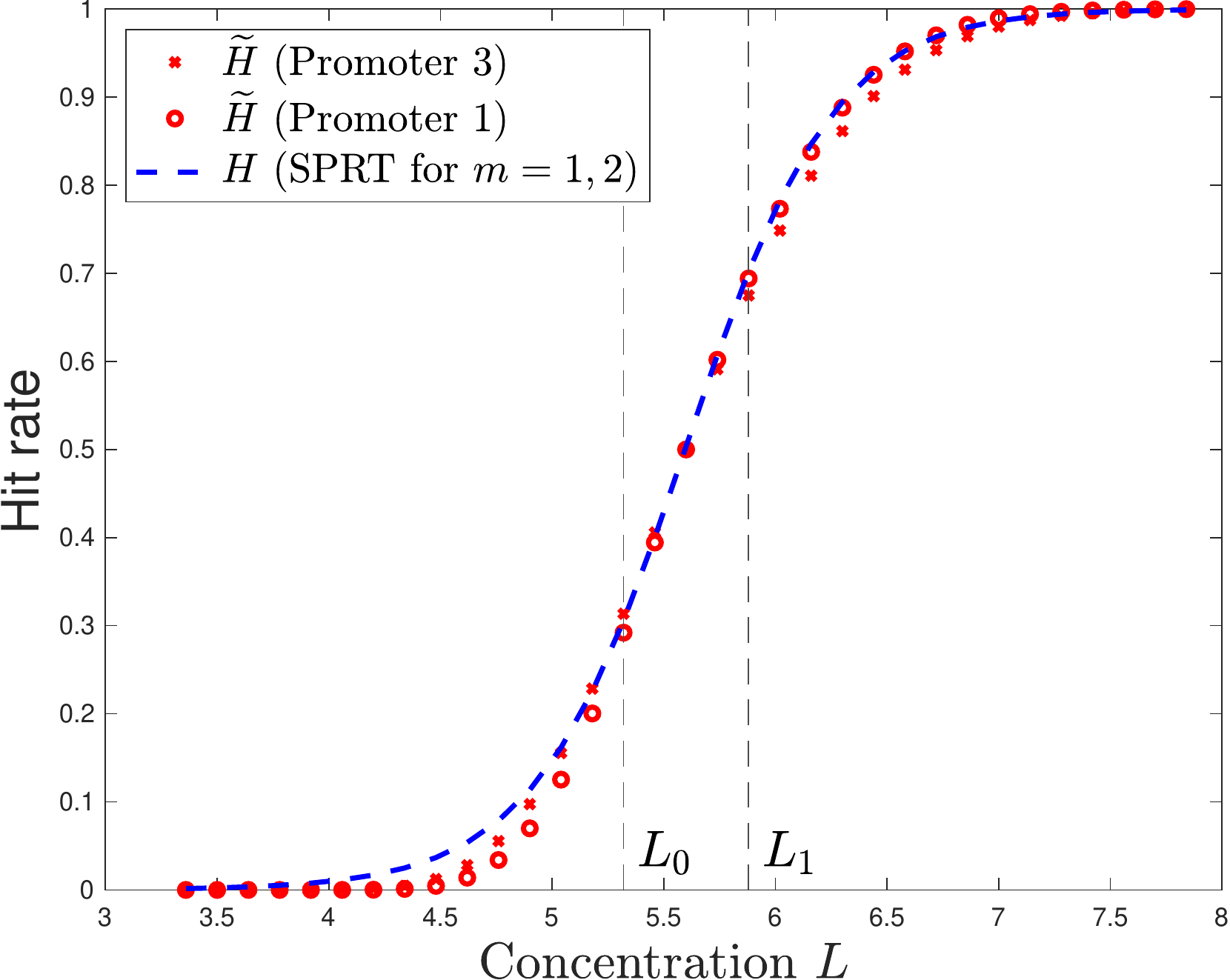} 
\caption{Comparing the hit rate $\widetilde{H}$ of Promoters 1 and 3, and that of SPRT ($H$). $K = 0.84$.}      
\label{fig:p1_p3_hr}
\end{figure}

\clearpage 

\section*{References}
\noindent{[\SciteCoxMiller]} D.R. Cox and H.D. Miller. \textsl{The Theory of Stochastic Processes}, Methuen \& Co,, London, 1965. \newline
\noindent{[\SciteChou]}  C.T. Chou. Designing Molecular Circuits for Approximate Maximum a Posteriori  Demodulation of Concentration Modulated Signals.
\emph{IEEE Transactions on Communications}, 67(8):5458--5473, July
  2019. \newline
\noindent{[\SciteSiggia]} E.~D. Siggia and M.~Vergassola. Decisions on the fly in cellular sensory
  systems. \emph{Proceedings of the National Academy of Sciences}, vol. 110,
  no.~39, pp. E3704--12, Sep. 2013. \newline

\noindent{[\SciteAhsendorf]} T. Ahsendorf, F. Wong, R. Eils, and J. Gunawardena. A framework for modelling gene regulation which
accommodates non-equilibrium mechanisms," \emph{BMC Biology}, vol. 12, no. 1, 2014. \newline

\noindent{[\SciteDesponds]} J. Desponds, M. Vergassola and A.M. Walczak, A mechanism for hunchback promoters to readout morphogenetic positional information in less than a minute. \emph{eLife}, 9:e49758.

\end{document}